\documentclass[10pt]{amsart}

\usepackage{amsfonts}
\usepackage{amssymb}
\usepackage{calrsfs}
\usepackage[mathscr]{eucal}
\usepackage{amsmath}
\usepackage{amsthm}
\usepackage[all]{xy}
\usepackage{refcheck}

\usepackage{hyperref}
\hypersetup{
    colorlinks=true,
    linkcolor=blue,
    citecolor=blue,
    filecolor=blue,
    urlcolor=blue
}

\newtheorem{thm}{Theorem}
\newtheorem{prop}[thm]{Proposition}
\newtheorem{cor}[thm]{Corollary}
\newtheorem{lem}[thm]{Lemma}
\theoremstyle{definition}
\newtheorem{defn}[thm]{Definition}
\newtheorem{rem}[thm]{Remark}
\newtheorem{ex}[thm]{Example}

\begin{document}

\title{Dual codes of product semi-linear codes}

\author{Luis Felipe Tapia Cuiti\~no and Andrea Luigi Tironi}

\date{November 7, 2013}

\address{
Departamento de Matem\'atica,
Universidad de Concepci\'on,
Casilla 160-C,
Concepci\'on, Chile}
\email{ltapiac@udec.cl,\ atironi@udec.cl}


\maketitle

\begin{abstract}
Let $\mathbb{F}_q$ be a finite field with $q$ elements and denote
by $\theta : \mathbb{F}_q\to\mathbb{F}_q$ an automorphism of
$\mathbb{F}_q$. In this paper, we deal with linear codes of
$\mathbb{F}_q^n$ invariant under a semi-linear map
$T:\mathbb{F}_q^n\to\mathbb{F}_q^n$ for some $n\geq 2$. In
particular, we study three kind of their dual codes, some
relations between them and we focus on codes which are products of
module skew codes in the non-commutative polynomial ring as a
subcase of linear codes invariant by a semi-linear map $T$. In
this setting we give also an algorithm for encoding, decoding and
detecting errors and we show a method to construct codes invariant
under a fixed $T$.
\end{abstract}

\section*{Introduction}

\noindent Recently there has been a lot of interest on algebraic codes in the setting of skew
polynomial rings which form an important family of non-commutative rings. Skew polynomials rings
have found applications in the construction of algebraic codes, where codes are defined as ideals (submodules) in the
quotient rings (modules) of skew polynomials rings.

The main motivation for considering these codes is that
polynomials in skew polynomial rings exhibit many factorizations
and hence there are many more ideals in a skew polynomial ring
than in the commutative case.

Furthermore, the research on codes in this setting has resulted in
the discovery of many new codes with better Hamming distance than
any previously known linear code with same parameters.

Inspired by the recent works \cite{BU1}, \cite{BU2} and \cite{BU3}, in $\S1$ we introduce the notion of $T$-codes, that is, linear codes invariant under a semi-linear transformation
$T$, and we characterize them from an algebraic point of view (see Theorem \ref{f-module inv}). In $\S 2$, as a consequence of Theorem \ref{rem Ore}
and Proposition \ref{prop Ore}, we introduce
the definition of \textit{product semi-linear $T$-codes}, a generalization of the module skew codes and a subcase of linear
codes invariant under a semi-linear transformation $T$ (see Definitions \ref{module skew codes} and \ref{product semilinear codes},
and Remark \ref{remark commutative case}). In particular, we show that in the commutative case any invariant code by means of an invertible linear transformation is isomorphic
as a vector space to a product of module codes (see Theorem \ref{rem Ore}).
In $\S 3$ we study the main properties of the Euclidean dual codes (Theorem \ref{Euclidean dual semilinear}, Proposition \ref{euclidean dual proposition} and Remark \ref{remark dual codes}),
the quasi-Euclidean and the Hermitian dual
codes (Definitions \ref{definition} and \ref{definition bis}, Theorems \ref{prop quasi euclidean} and \ref{T-theorem quasi-euclidean})
and the main relations among them (Theorem \ref{T-cor duals} and Corollaries \ref{cor quasi-euclidean and hermitian} and \ref{C+Ker B}).
Finally, in $\S 4$ we give an algorithm for encoding, decoding and detecting errors by a product semi-linear code (Algorithm $2$),
while in $\S 5$ we show a method to construct codes invariant under a semi-linear transformation (e.g., see Proposition \ref{remark fin} for the commutative case).

\section{Notation and background material}\label{section 2}

Let $\mathbb{F}_q$ be a finite field with $q$ elements and denote
by $\theta$ an automorphism of $\mathbb{F}_q$. Let us recall here
that if $q=p^s$ for some prime number $p$, then the map
$\tilde{\theta}:\mathbb{F}_q\to\mathbb{F}_q$ defined by
$\tilde{\theta}(a)=a^p$ is an automorphism on the field
$\mathbb{F}_q$ which fixes the subfield with $p$ elements. This
automorphism $\tilde{\theta}$ is called the {\em Frobenius
automorphism} and it has order $s$. Moreover, it is known
that the cyclic group it generates is the full group of
automorphisms of $\mathbb{F}_q$, i.e.
$\mathrm{Aut}(\mathbb{F}_q)=\langle\ \tilde{\theta}\ \rangle$. Therefore, any
$\theta\in\mathrm{Aut}(\mathbb{F}_q)$ is defined as
$\theta (a):=\tilde{\theta}^t(a)=a^{p^t},$ where $a\in\mathbb{F}_q$ and $t$ is an
integer such that $0\leq t\leq s$. Furthermore, when $\theta$ will be the identity automorphism
$id: \mathbb{F}_q\to\mathbb{F}_q$, we will write simply $\theta=id$.

If $n\geq 2$ is an integer, then we denote by $\mathbb{F}_q^n$ the vector space
$$\mathbb{F}_q^n:=\underbrace{\mathbb{F}_q\times
...\times\mathbb{F}_q}_{n-\mathrm{times}}.$$ It is well known that
a semi-linear map $T:\mathbb{F}_q^n\to\mathbb{F}_q^n$ is the
composition of an automorphism $\theta$ of $\mathbb{F}_q$ with an
$\mathbb{F}_q$-linear transformation $M$, i.e.
$(\vec{v})T:=(\vec{v})\Theta\circ M$, where $(v_1,...,v_n)\Theta
:=(\theta(v_1),...,\theta(v_n))$ and $M$ is an $n\times n$ matrix
with coordinates in $\mathbb{F}_q$. In this case we call $T$ a
\textit{$\theta$-semi-linear map}, or a
\textit{$\theta$-semi-linear transformation}.

For any $\vec{v}\in\mathbb{F}_q^n$ and any $T$ as above, let
$[\vec{v}]$ denote the $T$-cyclic subspace of $\mathbb{F}_q^n$
spanned by $\left\{\vec{v},(\vec{v})T,(\vec{v})T^2,... \right\}$.

Vector subspaces $\mathcal{C}_T\subset\mathbb{F}_q^n$ invariant by
a $\theta$-semi-linear transformation $T$ will be called here
\textit{semi-linear $T$-codes}, or \textit{$T$-codes} for simplicity.

The main result of \cite{J} allows us to decompose the vector
space $\mathbb{F}_q^n$ into a direct sum of very special vector
subspaces and to find a normal canonical form for any
$\theta$-semi-linear transformation.

\begin{defn}
We say that two $\theta$-semi-linear maps $T=\Theta\circ M$ and
$T'=\Theta\circ M'$ of $\mathbb{F}_q^n$ are $\theta$-similar if
there exists an invertible matrix $C$ such that $T=C^{-1}T'C$.
In this case we simply write $T\sim_\theta T'$.
\end{defn}

By choosing the basis of $\mathbb{F}_q^n$ to be the union of
appropriate bases
$$\left\{\vec{u}_i,(\vec{u}_i)T,(\vec{u}_i)T^2,...,(\vec{u}_i)T^{\dim[\vec{u}_i]-1} \right\}$$
of $T$-cyclic subspaces $[\vec{u}_i]$, $i=1,...,r$, it follows immediately the existence
of a normal canonical form for any $\theta$-semi-linear map $T$.

\begin{thm}[\cite{J}, Theorem 5]\label{semilinear}
Let $\theta$ and $T$ be an automorphism of $\mathbb{F}_q$ and a
$\theta$-semi-linear transformation on $\mathbb{F}_q^n$,
respectively. Then
$$\mathbb{F}_q^n=[\vec{u}_1]\oplus ...\oplus [\vec{u}_r],$$ for $T$-cyclic subspaces $[\vec{u}_i]$ satisfying
$\dim [\vec{u}_1]\geq\dim [\vec{u}_2]\geq ...\geq\dim
[\vec{u}_r]$. Moreover, if $T=\Theta\circ M$ then $$T\sim_{\theta}
\Theta\circ\mathrm{ diag}(M_1,...,M_r),$$ where
each $M_i$ is a matrix $n_i\times n_i$ of the following form
$$\left(
\begin{array}{c|ccc}
0 & 1 &  &  \\
\vdots &    & \ddots  &  \\
0 &   & & 1 \\
 \hline
a_{i,0} &  a_{i,1}  & \dots &  a_{i,n_i-1}
\end{array}\right)$$
with $n_i\geq 1$ and such that $\sum_{i=1}^{r}n_i=n$.
\end{thm}

\begin{rem}\label{semilinear-rem}
By Theorem \ref{semilinear}, we know that any $\theta$-semi-linear
transformation $T=\Theta\circ M$ is $\theta$-similar to
$$D:=\Theta\circ\mathrm{ diag}(M_1,...,M_r)=(\Theta\circ M_1,...,\Theta\circ M_r),$$ i.e. there exists an
invertible matrix
$$C:=\left(
\begin{array}{c}
C_1\\
\vdots \\
C_r \end{array}\right),\ \ \mathrm{where}\
C_i:=\left(
\begin{array}{c}
\vec{u}_i\\
(\vec{u}_i)T\\
\vdots \\
(\vec{u}_i)T^{n_i-1} \\
\end{array}\right) \ \mathrm{for\ every}\ i=1,...,r,$$
such that
$$T=C^{-1}DC=C^{-1}(\Theta\circ\mathrm{ diag}(M_1,...,M_r))C=C^{-1}(\Theta\circ M_1,...,\Theta\circ M_r)C,$$ where $n_i:=\dim[\vec{u}_i]$ for $i=1,...,r$
and each
$\Theta\circ M_i$ is the $\theta$-semi-linear transformation on
$\mathbb{F}_q^{n_i}$ such that
$\mathbb{F}_q^n=\mathbb{F}_q^{n_1}\times ...\times
\mathbb{F}_q^{n_r}$ with $\sum_{i=1}^{r} n_i = n$.
This gives a one -to-one correspondence between linear codes invariant under $T$ and linear codes invariant under $D$.
Therefore we can construct any semi-linear $D$-code $\mathcal{C}_D:=\mathcal{C}_TC^{-1}$ from a semi-linear $T$-code
$\mathcal{C}_T$, and vice versa.
\end{rem}

The proof of the below result is immediate.

\begin{lem}\label{lemma-euclidean}
We have the following two properties:
\begin{enumerate}
\item $\Theta\circ \overline{M}_{\theta}=\overline{M}\circ\Theta$, for any matrix $\overline{M}=[m_{ij}]$, where $\overline{M}_{\theta}:=[\theta(m_{ij})]$;
\item $(\vec{a}\ \Theta^{-1})\cdot\vec{b}=0\ \iff \ \vec{a}\cdot (\vec{b}\ \Theta)=0$,\ \ $\forall\ \vec{a},\vec{b}\in\mathbb{F}_q^n$.
\end{enumerate}
\end{lem}

\begin{rem}\label{semilinear-rem-bis}
Let $T=\Theta\circ M$ be a $\theta$-semi-linear transformation. If
$M$ is an $n\times n$ matrix with coordinates in
$\mathbb{F}_q^{\theta}\subseteq\mathbb{F}_q$, the subfields of
$\mathbb{F}_q$ fixed by $\theta$, then $M$ admits a rational
normal form (obtained by the Magma~\cite{M} command {\tt RationalForm(M)}), i.e. there exists an invertible matrix $C$ with
coordinates in $\mathbb{F}_q^{\theta}$ such that $M=C^{-1}M'C$,
where $M':=\mathrm{ diag}(M_1,...,M_k)$ and each $M_i$ is a
$n_i\times n_i$ matrix as in Theorem \ref{semilinear} defined over
$\mathbb{F}_q^{\theta}$. Thus we have
$$CTC^{-1}=C(\Theta\circ M) C^{-1}=\Theta\circ CMC^{-1} =\Theta\circ M'=D$$ and in this case it is easy to find a matrix $C$
which transforms a $T$-code into a $D$-code, and vice versa. Typical examples of this situation are the
skew quasi-cyclic codes, where the matrix $M$ is a permutation
matrix $P$ such that $P=P_{\theta}=P_{\theta^{-1}}$.
\end{rem}

Consider the ring structure defined on the following set:
$$R:=\mathbb{F}_q[X;\theta]=\left\{ a_sX^s+...+a_1X+a_0\ | \ a_i\in\mathbb{F}_q\ \mathrm{and}\ s\in\mathbb{N}\right\}.$$
The addition in $R$ is defined to be the usual addition of
polynomials and the multiplication is defined by the basic rule
$X\cdot a = \theta(a)X$ for any $a\in\mathbb{F}_q$ and extended to
all elements of $R$ by associativity and distributivity. The ring
$R$ is known as skew polynomial ring and its elements are skew
polynomials. Moreover, it is a right Euclidean ring whose
left ideals are principals.

\medskip

From now on, together with the same notation as above, we will always assume the following

\bigskip

\noindent {\bf Hypothesis $(*)$ :} \quad $T=\Theta\circ M$ is a fixed
$\theta$-semi-linear transformation of $\mathbb{F}_q^n$ which is $\theta$-similar to
$D:=\Theta\circ\mathrm{ diag}(M_1,...,M_r)$ by a matrix $C$
and $f_j:=(-1)^{n_j}(X^{n_j}-\sum_{i=0}^{n_j-1}a_{j,i}X^i) \in R$
is the characteristic polynomial of $M_j$ with $a_{j,0}\neq 0$, where the coefficients
$a_{j,i}$ are given by Theorem \ref{semilinear} for every $j=1,...,r$ and $i=0,...,n_j-1$.

\bigskip

Denote by
$\pi_j:\mathbb{F}_q^{n_j}\to R/Rf_j$ the linear transformation
which sends a vector
$\vec{c}_j=(c_0,...,c_{n_j-1})\in\mathbb{F}_q^{n_j}$ to the
polynomial class $c_j(X)=\sum_{i=0}^{n_j-1}c_iX^i$ of $R/Rf_j$.

Moreover, consider the linear map
$$\pi :\ \mathbb{F}_q^{n_1}\times ...\times \mathbb{F}_q^{n_r}\to R_n:=R/Rf_1\times ...\times R/Rf_r , $$
where $\pi=(\pi_1,...,\pi_r)$ and the linear transformation
$\pi_j: \mathbb{F}_q^{n_j}\to R/Rf_j$ is defined as above for each
$j=1,...,r$.

Let $\mathcal{C}\subseteq\mathbb{F}_q^n$ be a linear code and define the linear code
$$\mathcal{C}C^{-1}:=\{\ \vec{c}\ C^{-1}\in
\mathbb{F}_q^n\ |\ \vec{c}\in\mathcal{C}\ \}.$$

We can obtain now the following characterization of any $T$-code
in $\mathbb{F}_q^n$.

\begin{thm}\label{f-module inv}
With the same notation as in $(*)$, let
$\mathcal{C}\subseteq\mathbb{F}_q^n$ be a linear code and put $\mathcal{C}':=\mathcal{C}C^{-1}$. Then

$\mathcal{C}$ is a $T$-code $\iff$ $\mathcal{C}'$ is a linear code invariant under $D$
$\iff$ $\pi (\mathcal{C}')$ is a left $R$-submodule of $R_n$.

\end{thm}

\noindent\textit{Proof}. From Remark \ref{semilinear-rem}, we know
that any $T$-code can be written as $\mathcal{C}'C$, where
$\mathcal{C}'$ is a linear code invariant by
$D$, and vice versa. So it is sufficient to
show that a linear code $\mathcal{C}'$ is invariant under $D$ if and
only if $\pi (\mathcal{C}')$ is a left $R$-submodule of $R_n$.
Let $\mathcal{C}'$ be a linear code invariant by $D$. Note that
$\pi (\mathcal{C}')$ is an abelian group with respect to the sum.
Moreover, observe that
$X\cdot \pi (\vec{v}) = \pi (\vec{v}D)\in\pi(\mathcal{C}')$ for any $\vec{v}\in\mathcal{C}'$.
By an inductive argument and linearity, this implies that $g\cdot \pi (\vec{v})\in\pi(\mathcal{C}')$ for any $g\in R$, that is, $\pi (\mathcal{C}')$ is an $R$-submodule of
$R_n$. On the other hand, let $\pi (\mathcal{C}')$ be an $R$-submodule of
$R_n$. Then $\mathcal{C}'=\pi^{-1}(\pi(\mathcal{C}'))$ is a vector
subspace of $\mathbb{F}_q^n$ and for every $\vec{c}\in
\mathcal{C}'$ we have
$\vec{c}D=\pi^{-1}(X\cdot\pi (\vec{c}))\in\pi^{-1}(\pi (\mathcal{C}'))=\mathcal{C}',$
since $X\cdot\pi (\vec{c})\in \pi (\mathcal{C}')$. \hfill
$\square$

\medskip

\begin{rem}
If $T=\Theta\circ M_1$, where $M_1$ is a matrix as in Theorem
\ref{semilinear} with $a_{1,0}\neq 0$, then $C$ in $(*)$ is the
identity matrix and the above result becomes a geometric
characterization of the module $\theta$-codes (see \cite[Definition 1
and Proposition 1]{BU1}) associated to the polynomial
$f_1:=(-1)^{n_1}(X^{n_1}-a_{1,n_1-1}X^{n_1-1}-...-a_{1,0})$. Moreover, if
$\theta=id$, then Theorem \ref{f-module inv} generalizes \cite[(2.1)]{LL}.

\end{rem}

\begin{ex}
In $\mathbb{F}_{4}^6$, where
$\mathbb{F}_{4}=\mathbb{F}_{2}[\alpha]$ with
$\alpha^2+\alpha+1=0$, consider the matrix
$$D=\left(
\begin{array}{c|c}
E & O \\
\hline
O & E
\end{array}\right),\ \ \mathrm{where}\ E=\left(
\begin{array}{ccc}
0 & 1 & 0 \\
0 & 0 & 1 \\
1 & 0 & 0 \\
\end{array}\right)\ \mathrm{and} \
O=\left(
\begin{array}{ccc}
0 & 0 & 0 \\
0 & 0 & 0 \\
0 & 0 & 0 \\
\end{array}\right),
$$
and the semi-linear transformation $\Theta\circ D$. The code $\mathcal{C}=\langle\ (1,1,1,1,1,1)\ \rangle$ is invariant by
$\Theta\circ D$, $\mathcal{C}\cong\langle\ (1,1,1,0,0,0)\ \rangle=\langle\ (1,1,1)\ \rangle\times
\langle\ (0,0,0)\ \rangle$, but $\mathcal{C}\neq\mathcal{C}_1\times\mathcal{C}_2$
for any $\theta$-code $\mathcal{C}_i\subseteq\mathbb{F}_{4}^3$
invariant by $\Theta\circ E$ for $i=1,2$.
\end{ex}

\begin{rem}
Remark \ref{semilinear-rem} and Theorem \ref{f-module inv} say that there is a one-to-one correspondence
between $T$-codes and left $R$-submodules of $R_n$.
\end{rem}

\begin{rem}
In the commutative case, i.e. $\theta=id$, the Chinese
Remainder Theorem says that if $(f_1),...,(f_k)$ are ideals of $R$
which are pairwise coprime, that is $(f_i)+(f_j)=R$ for all $i\neq
j$, then $I:=(f_1)\cap ...\cap (f_k)=(f_1)\cdot ...\cdot (f_k)$
and the quotient ring $R/I$ is isomorphic to the product ring
$R/(f_1)\times ... \times R/(f_k)$ via the isomorphism $f:R/I\to
R/(f_1)\times ... \times R/(f_k)$ such that
$f(a+I):=(a+(f_1),...,a+(f_k))$.

In the non-commutative case there exists an analogous of the above
result. When $\theta\neq id$, if $Rf_1,...,Rf_k$ are pairwise
coprime two-sided ideals of $R$, then
$$R/(Rf_1\cap ...\cap Rf_k)\cong R/Rf_1\times ... \times R/Rf_k $$
as $R$-modules and $I:=Rf_1\cap ...\cap Rf_k$ can be replaced by a
sum over all orderings of $Rf_1,...,Rf_k$ of their product (or
just a sum over enough orderings, using inductively that $J\cap
K=JK+KJ$ for coprimes ideals $J,K$). In both situations, we have a
method to find all the $R$-submodules of $R/Rf_1\times ... \times
R/Rf_k$ via $R/I$.
\end{rem}

\begin{cor}
Assume that $\theta=id$ and write $T=M$ with $M=C^{-1}M_1C$, where $M_1$ is as in $(*)$.
If $f_1=\Pi_{i=1}^{s}p_i^{e_i}$ is the factorization of the polynomial $f_1$ in $\mathbb{F}_q[x]$, where $p_1,...,p_s$ are distinct monic irreducible polynomials and
$e_i\in\mathbb{N}-\{0\}$ for all $i=1,...,s$, then there exist $\Pi_{i=1}^{s}(e_i+1)$ $T$-codes $\mathcal{C}\subseteq\mathbb{F}_q^{\deg f_1}$.
\end{cor}

\noindent\textit{Proof}. The statement follows from Remark \ref{semilinear-rem}
and Theorem \ref{f-module inv} by counting the number of all monic divisors of $f_1$.
\hfill $\square$

\medskip

By using the computer algebra system Magma \cite{M}, the following MAGMA program enables us to find 
all the right divisors of any polynomial $f\in\mathbb{F}_a[X;\theta] :$

\begin{verbatim}
F<w>:=GF(a);
E:=[x : x in F | x ne 0];
RightDivisors := function(qq,g) 
 R<x>:=TwistedPolynomials(F:q:=qq); 
 f:=R!g;
 n:=Degree(f); 
 S:=CartesianProduct(E,CartesianPower(F,n-1)); 
 dd:=[]; 
  for ss in S do
   ll:=[ss[1]] cat [p : p in ss[2]]; 
   q,r:=Quotrem(f,R!ll); 
    if r eq R![0] then 
    dd := dd cat [[q,R!ll]]; 
    end if;
  end for;
 return dd; 
end function;
\end{verbatim}

\medskip

Finally, we have the following two results.

\begin{thm}\label{rem Ore}
Suppose that $\theta =id$. Then any $R$-submodule $S$ of
$R_n=R/Rf_1\times ...\times R/Rf_r$ is $R$-isomorphic to a
product $S_1\times ...\times S_r$, where each $S_j$ is an $R$-submodule
of $R/Rf_j$ for every $j=1,...,r$. In particular, any $D$-code
$\mathcal{C}_D\subseteq\mathbb{F}_q^n$ with $D=\mathrm{ diag}(M_1,...,M_r)$ is isomorphic to a product code
$\mathcal{C}_1\times\dots\times\mathcal{C}_r\subseteq\mathbb{F}_q^{n_1}\times
...\times \mathbb{F}_q^{n_r}$ as a vector
subspace of $\mathbb{F}_q^n=\mathbb{F}_q^{n_1}\times ...\times
\mathbb{F}_q^{n_r}$, i.e. $\mathcal{C}_D=(\mathcal{C}_1\times\dots\times\mathcal{C}_r)\widehat{C}$
for some invertible matrix $\widehat{C}$, where each
$\mathcal{C}_i\subseteq\mathbb{F}_q^{n_i}$ is a linear code
invariant by $M_i$, $M_i$ being the $n_i\times n_i$
matrix of {\em Theorem \ref{semilinear}}.
\end{thm}

\noindent\textit{Proof}. It is sufficient to prove the first part
of the statement for $r\geq 2$, since the second one follows
immediately from this by putting $n_i:=\deg f_i$ for $i=1,...,r$.

If each polynomial $f_j\in R$ is written as a product
$F_{j1}^{a_{j1}}\cdots F_{jt_j}^{a_{jt_j}}$ of distinct
irreducible polynomials $F_{jk}$ for some integers $a_{ji}\geq 1$, then by
the Chinese Reminder Theorem we can obtain via isomorphisms a
decomposition $A$ of $R_n=R/Rf_1\times ...\times R/Rf_r$ such that
$$A:=(R/RF_{11}^{a_{11}}\times ... \times R/RF_{1t_1}^{a_{1t_1}})\times ... \times (R/RF_{r1}^{a_{r1}}\times ... \times R/RF_{rt_r}^{a_{rt_r}})\cong R_n .$$

\noindent Let $S$ be an $R$-submodule of $R_n$. Then, up
to isomorphisms, $S$ corresponds to an $R$-submodule $S'$ of $A$.
Thus we have to prove only that every $R$-submodule $S'$ of
$A$ is isomorphic to a product $S_{11}\times ... \times S_{rt_r}\subseteq A$
of $R$-submodules $S_{ij_i}\subseteq R/RF_{ij_i}^{a_{ij_i}}$ for every $i=1,...,r$ and $j_i=1,...,t_i$.

So, let $W$ be an $R$-submodule of $A$. Then $W$ is $R$-isomorphic to
a direct sum $Rg_1\oplus\cdots\oplus Rg_k$ of non-zero
distinct cyclic $R$-submodules $Rg_i$ of $A$ with $g_i\in A$ for
$i=1,...,k$. Consider the surjective $R$-homomorphism $\pi_i: R\to Rg_i$ and
note that $Rg_i\cong R/(\mathrm{Ker}\ \pi_i)$ for any $i=1,...,k$. Since $R$ is a principal ideal domain, we see that
$\mathrm{Ker}\ \pi_i=(p_i)$ for some $p_i\in R$. Let $F$ be the product
$F_{11}^{a_{11}}\cdot ... \cdot F_{st_s}^{a_{st_s}}$
of all distinct polynomials with the respective maximum powers which appear in the decompositions $F_{j1}^{a_{j1}}\cdots F_{jt_j}^{a_{jt_j}}$ of the
polynomials $f_j$. Then we deduce that $F\in\mathrm{Ker}\ \pi_i=(p_i)$, i.e.
for every $i=1,...,k$ there exists a polynomial $q_i$ such that $F=q_ip_i$. This implies that $p_i=F_{11}^{c_{11}}\cdot ... \cdot F_{st_s}^{c_{st_s}}$
for some integers $c_{jt_j}$ such that $0\leq c_{jt_j}\leq a_{jt_j}$ for every $i=1,...,k$ and $j=1,...,s$.
So we conclude that
$$Rg_i\cong R/(p_i)=R/(F_{11}^{c_{11}}\cdot ... \cdot F_{st_s}^{c_{st_s}})\cong
R/F_{11}^{c_{11}}\times ... \times R/F_{st_s}^{c_{st_s}}\subseteq A,$$
i.e. $Rg_i\cong RF_{11}^{a_{11}-c_{11}}/F_{11}^{a_{11}}\times ... \times RF_{st_s}^{a_{st_s}-c_{st_s}}/F_{st_s}^{a_{st_s}}\cong S_{11}\times ... \times S_{rt_r}\subseteq A$, where
$\{0\}\subseteq S_{ij_i}\subseteq R/RF_{ij_i}^{a_{ij_i}}$ is an $R$-submodule for every $i=1,...,r$ and $j_i=1,...,t_i$.
\hfill $\square$

\begin{prop}\label{prop Ore}
Suppose that $\theta\neq id$. If $R_n=R/Rf_1\times ...\times R/Rf_r$ with
$Rf_1,...,Rf_r$ pairwise coprime two-sided ideals of $R$, then any $R$-submodule $S$ of
$R_n=R/Rf_1\times ...\times R/Rf_r$ is $R$-isomorphic to a
product $S_1\times ...\times S_r$, where each $S_j$ is an $R$-submodule
of $R/Rf_j$ for every $j=1,...,r$. In particular, under the above hypothesis, any $D$-code
$\mathcal{C}_D\subseteq\mathbb{F}_q^n$ is isomorphic to a product code
$\mathcal{C}_1\times\dots\times\mathcal{C}_r\subseteq\mathbb{F}_q^{n_1}\times
...\times \mathbb{F}_q^{n_r}$ as a vector
subspace of $\mathbb{F}_q^n=\mathbb{F}_q^{n_1}\times ...\times
\mathbb{F}_q^{n_r}$, i.e. $\mathcal{C}_D=(\mathcal{C}_1\times\dots\times\mathcal{C}_r)\widehat{C}$
for some invertible matrix $\widehat{C}$, where each
$\mathcal{C}_i\subseteq\mathbb{F}_q^{n_i}$ is a linear code
invariant by $\Theta\circ M_i$, $M_i$ being the $n_i\times n_i$
matrix of {\em Theorem \ref{semilinear}}.
\end{prop}

\noindent\textit{Proof}. Let $S$ be an $R$-submodule of $R_n$. Then $S$ is $R$-isomorphic to
a direct sum $Rg_1\oplus\cdots\oplus Rg_k$ of non-zero
distinct cyclic $R$-submodules $Rg_i$ of $R_n$ with $g_i\in R_n$ for
$i=1,...,k$. Write $g_i=(g_{i1},...,g_{ir})$ and consider the polynomial $F:=f_{1}\cdot ... \cdot f_{r}$.
Denote by $F_{h}$ the product $F$ without the factor $f_{h}$. Then we get
$$F_{h}g_i=(0,...,0,F_{h}g_{ih},0,...,0).$$ Since the (right) $gcd(f_h,F_h)=1$, we know that there exist two polynomials
$a,b\in R$ such that $af_h+bF_h=1$. Hence $bF_{h}g_i=(0,...,0,g_{ih},0,...,0)$ for every $i=1,...,k$ and $h=1,...,r$.
Therefore we have
$$Rg_i=R(g_{i1},0,...,0)\oplus ...\oplus R(0,...,0,g_{ir})\cong (Rg_{i1},...,Rg_{ir})$$ for every $i=1,...,k$, i.e.
$S\cong Rg_1\oplus\cdots\oplus Rg_k \cong (S_1,...,S_r)$ for some
$R$-submodules $S_j\subset R/Rf_j$, where $j=1,...,r$.
\hfill $\square$

\section{Product semi-linear codes}

Let us recall here the following

\begin{defn}[see \cite{BU1}]\label{module skew codes}
An \textit{$f_j$-module $\theta$-code} (or simply a \textit{module
$\theta$-code}) $\mathcal{C}_j$ is a linear code in
$\mathbb{F}_q^{n_j}$ which corresponds via
$\pi_j:\mathbb{F}_q^{n_j}\to R/Rf_j$ to a left $R$-submodule
$Rg_j/Rf_j\subset R/Rf_j$ in the basis $1,X,...,X^{n_j-1}$, where
$g_j$ is a right divisor of $f_j$ in $R$. The length of the code
$\mathcal{C}_j$ is $n_j=\deg(f_j)$ and its dimension is
$k_j=\deg(f_j)-\deg(g_j)$. For simplicity, we will denote this
code $\mathcal{C}_j=(g_j)_{n_j,q}^{k_j,\theta}$ and
when there will not be any confusion, we will call an
$f_j$-module $\theta$-code simply a module $\theta$-code.
\end{defn}

\begin{rem}
When $\theta=id$, Definition \ref{module skew codes} coincides with the one of
a generalized cyclic code with respect to a polynomial $f_j$ (see \cite{LL} and \cite{MacWS}).
\end{rem}

Therefore, from Theorem \ref{rem Ore}, Proposition \ref{prop Ore} and Definition \ref{module skew codes}, it follows naturally the
below definition.

\begin{defn}\label{product semilinear codes}
Let $\mathcal{C}_T\subseteq\mathbb{F}_q^n$ be a semi-linear $T$-code
invariant by a $\theta$-semi-linear map $T$ as in $(*)$. We say that
$\mathcal{C}_T$ is a \textit{product semi-linear $T$-code}, or
\textit{a product $T$-code}, if $\mathcal{C}_T=(\mathcal{C}_1\times
...\times\mathcal{C}_r)C\subseteq\mathbb{F}_q^{n_1}\times
...\times \mathbb{F}_q^{n_r}$, where
any $\mathcal{C}_j\subseteq\mathbb{F}_q^{n_j}$ is an $f_j$-module
$\theta$-codes with respect to $\Theta\circ M_j$ and
$f_j=(-1)^{n_j}(X^{n_j}-\sum_{k=0}^{n_j-1}a_{j,k}X^{k})$
is as in $(*)$ for every $j=1,...,r$ and $n=\sum_{j=1}^{r}n_j$.
\end{defn}

\begin{rem}
When $C$ is the identity matrix and $r=1$, then Definition \ref{product semilinear codes} is nothing else that the definition of an $f_1$-module $\theta$-code.
\end{rem}

\begin{rem}\label{remark commutative case}
When either $\theta =id$, or $\theta\neq id$ and $R_n=R/Rf_1\times ...\times R/Rf_r$ with
$Rf_1,...,Rf_r$ pairwise coprime two-sided ideals of $R$, Theorem \ref{rem Ore} and Proposition \ref{prop Ore} show that any $T$-code $\mathcal{C}_T$ is isomorphic
to a product $T$-code as vector spaces, i.e. for any $T$-code $\mathcal{C}_T\subseteq\mathbb{F}_q^n=\mathbb{F}_q^{n_1}\times
...\times \mathbb{F}_q^{n_r}$
there exists an invertible matrix $C'$ such that $\mathcal{C}_T=(\mathcal{C}_1\times
...\times\mathcal{C}_r)CC'$ for some $T$-product code $(\mathcal{C}_1\times
...\times\mathcal{C}_r)C\subseteq\mathbb{F}_q^{n_1}\times
...\times \mathbb{F}_q^{n_r}$.
\end{rem}

From Definition \ref{product semilinear codes} we deduce that a
generator matrix of a product semi-linear code
$\mathcal{C}_T=(\mathcal{C}_1\times
...\times\mathcal{C}_r)C$ is given by

$$\left(\begin{array}{cccc}
G_1 &  &  &  \\ [2pt]
 & G_2 &  &  \\  [2pt]
 &  & \ddots &  \\  [2pt]
 &  &  & G_r \\  [2pt]
\end{array}\right)\cdot C,$$
where $k_i:=\dim\mathcal{C}_i$, $\sum_{i=1}^{r}k_i=\dim
\mathcal{C}_T$ and each block
$$G_i:=\left(
\begin{array}{c}
\vec{g}_i  \\
(\vec{g}_i)(\Theta\circ M_i) \\
\vdots  \\
(\vec{g}_i)(\Theta\circ M_i)^{k_i-1}
\end{array}\right)$$
is a $k_i\times n_i$ generator matrix of the module $\theta$-code
$\mathcal{C}_i=(g_i)_{n_i,\theta}^{k_i}$, where
$\vec{g}_i=\pi_i^{-1}(g_i)$ and $\pi_i:\mathbb{F}_q^{n_i}\to
R/Rf_i$ for every $i=1,...,r$.

\medskip

\begin{defn}
A linear code $\mathcal{C}$ is a code of type $[n,k]_q$ if
$\mathcal{C}\subseteq\mathbb{F}_q^n$ and
$\dim_{\mathbb{F}_q}\mathcal{C}=k$.
\end{defn}

The following result gives in the commutative case a necessary and sufficient condition
for the existence of $T$-codes $\mathcal{C}_T$ of type $[n,k]_q$.

\begin{prop}
Suppose that $\theta=id$ and let $T=M$ be a linear transformation
over $\mathbb{F}_q^n$ as in $(*)$. Let
$\mathbb{F}_q^n=\mathbb{F}_q^{N_1}\times ...\times
\mathbb{F}_q^{N_s}$ be a decomposition of $\mathbb{F}_q^n$ as in
the proof of {\em Theorem \ref{rem Ore}} and denote by $\pi$ the
corresponding isomorphism
$$\pi : \mathbb{F}_q^n=\mathbb{F}_q^{N_1}\times ...\times \mathbb{F}_q^{N_s}\to
R/RF_1^{\alpha_1}\times ...\times R/RF_s^{\alpha_s},$$ where
$\pi=(\pi_1,...,\pi_s)$ and $\pi_j: \mathbb{F}_q^{N_j}\to R/RF_j$
are the usual isomorphisms and the $F_j$'s are irreducible (not
necessarily distinct) polynomials on $R$ such that
$N_j=\alpha_j\deg F_j\geq 1$ for $j=1,...,s$. Then

\medskip

$\exists$ a $T$-code of type $[n,k]_q$ $\iff$
$k=\sum_{i=1}^{s}a_i\deg F_{i}$, where $0\leq a_i\leq \alpha_i$.

\end{prop}

\noindent\textit{Proof}. Note that for every $i=1,...,s$ an $R$-submodule of
$R/RF_i^{\alpha_i}$ is of type $RF_i^h/RF_i^{\alpha_i}\cong
R/RF_i^{\alpha_i-h}$ for some integer $h$ such that $0\leq h\leq
\alpha_i$. Moreover, observe that by
Remark \ref{semilinear-rem} the set of the $T$-codes
$\mathcal{C}_T$ is in one-to-one correspondence with the set of
linear codes $\mathcal{C}_D$ invariant by the linear
transformation $D:=\mathrm{ diag}(M_1,...,M_r)$ of type $[n,k]_q$.
Let $\mathcal{C}_T\subset\mathbb{F}_q^n$ be a $T$-code of type
$[n,k]_q$. Then $\mathcal{C}_TC^{-1}$ is a linear code
$\mathcal{C}_D$ invariant by the linear transformation
$D:=\mathrm{ diag}(M_1,...,M_r)$. With the same notation as in the
statement, $\pi(\mathcal{C}_D)$ is an $R$-submodule of
$R/RF_1^{\alpha_1}\times ...\times R/RF_s^{\alpha_s}$. Since by
Theorem \ref{rem Ore} every $R$-submodule of
$R/RF_1^{\alpha_1}\times ...\times R/RF_s^{\alpha_s}$ is
isomorphic to $I_1\times ...\times I_s$ with $I_j$ an $R$-submdule
of $R/RF_{j}^{\alpha_j}$ for every $j=1,...,s$, we conclude that
$k:=\dim \mathcal{C}_T=\dim \mathcal{C}_D=\sum_{i=1}^{s}a_i\deg F_{i},$
where $0\leq a_i\leq \alpha_i$. On the other hand,
assume that $k=\sum_{i=1}^{s}a_i\deg F_{i}$ with $0\leq a_i\leq
\alpha_i$. Then the product code
$$\pi_1^{-1}(RF_1^{\alpha_1-a_1}/RF_1^{\alpha_1})\times
...\times \pi_s^{-1}(RF_s^{\alpha_s-a_s}/RF_1^{\alpha_s})=$$
$$=\pi^{-1}(RF_1^{\alpha_1-a_1}/RF_1^{\alpha_1}\times ...\times
RF_s^{\alpha_s-a_s}/RF_1^{\alpha_s})$$ is a $T$-code of type
$[n,k]_q$. \hfill $\square$

\section{Dual codes of product $T$-codes}

In this section we study three kind of dual codes of product
semi-linear $T$-codes and some main relations between them.

\subsection{Euclidean duals}

In \cite{BU1} the authors prove that the Euclidean dual code of a
module $\theta$-code is a module $\theta$-code if and only if it
is a $\theta$-constacyclic code. Moreover, they establish that a
module $\theta$-code which is not $\theta$-constacyclic code is a
shortened $\theta$-constacyclic code and that its Euclidean dual
is a punctured $\theta$-constacyclic code. This enables them to
give a form of the parity-check matrix for module $\theta$-codes.

Let us only observe here that there exists an alternative method
to find a parity-check matrix for any module $\theta$-code.

\begin{prop}\label{pc matrix}
Let $\mathcal{C}_j=(g_j)_{n_j,\theta}^{k_j}\subseteq\mathbb{F}_q^{n_j}$ be a module
$\theta$-code. For any integer $i$ such that $0\leq i\leq k_j-1$,
write in $R$
$$X^{n_j-k_j+i}=q_ig_j+r_i,\ \mathrm{with\ } 0\leq\deg r_i<n_j-k_j.$$
Denote by $S$ the following matrix
$$S:=\left(
\begin{array}{c}
\rho_{n_j-k_j}(\pi_j^{-1}(r_0)) \\
\rho_{n_j-k_j}(\pi_j^{-1}(r_1)) \\
\vdots \\
\rho_{n_j-k_j}(\pi_j^{-1}(r_{k_j-1}))
\end{array}\right) ,$$ where $\pi_j:\mathbb{F}_q^{n_j}\to R/Rf_j$ and $\rho_{n_j-k_j}$ is the projection map onto the first $n_j-k_j$ coordinates, i.e.
$$\rho_{n_j-k_j}(v_1,...,v_{n_j-k_j},v_{n_j-k_j+1},...,v_{n_j}):=(v_1,...,v_{n_j-k_j}).$$
Then a generator matrix $G_j$ of $\mathcal{C}_j$ is
$$G_j:=\left(
\begin{array}{c|c}
-S & I_{k_j}
\end{array}\right)$$
and a parity check matrix $H_j$ is given by
$$H_j:=\left(
\begin{array}{c|c}
I_{n_j-k_j} & S_t
\end{array}\right), $$ where $I_{n_j-k_j}$ is the $(n_j-k_j)\times (n_j-k_j)$ identity matrix and
$S_t$ is the transpose matrix of $S$.
\end{prop}

\noindent\textit{Proof}. Since $\deg r_i<n_j-k_j$, note that
$\pi_j^{-1}(X^{n_j-k_j+i}-r_i)\in \mathcal{C}_j$
are linearly independent for $0\leq i\leq k_j-1$. Thus
$\left(
\begin{array}{c|c}
-S & I_{k_j}
\end{array}\right)$
is a generator matrix $G_j$ for the code $\mathcal{C}_j$. Moreover, since
$(\mathcal{C}_j^{\perp})^{\perp}=\mathcal{C}_j$, we see that
the matrix $H_j:=\left(
\begin{array}{c|c}
I_{n_j-k_j} & S_t
\end{array}\right)$ as in the statement is a parity check matrix for
$\mathcal{C}_j$. \hfill $\square$

\bigskip

The following MAGMA program enables us to find all the polynomials 
$r_i$ of Proposition~\ref{pc matrix} in $\mathbb{F}_a[X;\theta] :$

\begin{verbatim}
F<w>:=GF(a); 
PcMatrix:=function(qq,g,n) 
 R<x>:=TwistedPolynomials(F:q:=qq); 
 g:=R!g; 
 d:=Degree(g); 
 ll:=[]; 
  for i in [0.. n-d-1] do 
   c,b:=Quotrem(R![0,1]^(d+i),g); 
   ll:=ll cat [b]; 
  end for; 
 return ll; 
end function;
\end{verbatim}

\begin{rem}
Proposition \ref{pc matrix} works also for any module $(\theta,\delta)$-code
(see \cite[Definition 1]{BU2}), where $\delta :\mathbb{F}_q\to\mathbb{F}_q$ is a derivation, and it
allows us to obtain directly a generator and a parity-check matrix in standard form for any module $(\theta,\delta)$-code.
\end{rem}

\begin{thm}\label{Euclidean dual semilinear}
Let $\mathcal{C}_T=(\mathcal{C}_1\times ...\times\mathcal{C}_r)\widehat{C}\subseteq\mathbb{F}_q^n$ be a linear
code, $\mathcal{C}_i\subseteq\mathbb{F}_q^{n_i}$ being a linear code and $\mathbb{F}_q^n=\mathbb{F}_q^{n_1}\times ...\times\mathbb{F}_q^{n_r}$.
If $\widehat{C}$ is an invertible matrix, then
$${\mathcal{C}_T}^{\perp}=(\mathcal{C}_1^{\perp}\times ...\times \mathcal{C}_r^{\perp})\widehat{C}_{t}^{-1},$$
where $\widehat{C}_t$ is the transpose matrix of $\widehat{C}$ and
$\mathcal{C}_i^{\perp}\subseteq \mathbb{F}_q^{n_i}$ is the Euclidean dual code of
$\mathcal{C}_i$ for every $i=1,...,r$. Furthermore, a parity check
matrix of $\mathcal{C}_T$ is
$$\left(\begin{array}{ccc}
H_1 &  &    \\
 &  \ddots &  \\
 &  & H_r \\
\end{array}\right)\cdot \widehat{C}_t^{-1}$$
where $h_i:=\dim\mathcal{C}_i^{\perp}$, $\sum_{i=1}^{r}h_i=\dim
\mathcal{C}_T^{\perp}$ and $H_i$ is the $h_i\times n_i$ parity
check matrix of $\mathcal{C}_i$ given by {\em Proposition \ref{pc
matrix}} for every $i=1,...,r$.
\end{thm}

\noindent\textit{Proof}. Put $\mathcal{C}:=(\mathcal{C}_1^{\perp}\times
...\times \mathcal{C}_r^{\perp})\widehat{C}_{t}^{-1}$ and note
that
$$\dim\mathcal{C} = \dim (\mathcal{C}_1^{\perp}\times ...\times \mathcal{C}_r^{\perp})
=\sum_{i=1}^{r}\dim
\mathcal{C}_i^{\perp}=\sum_{i=1}^{r}(m_i-\dim\mathcal{C}_i)=$$
$$=\sum_{i=1}^{r}m_i-\sum_{i=1}^{r}\dim (\mathcal{C}_i)=n-\dim \mathcal{C}_T=\dim {\mathcal{C}_T}^{\perp}.$$

Let $\vec{v}\in \mathcal{C}$. Since $\mathcal{C}=(\mathcal{C}_1^{\perp}\times
...\times \mathcal{C}_r^{\perp})\widehat{C}_{t}^{-1}$, we deduce
that $\vec{v}=\vec{w}\widehat{C}_{t}^{-1}$ for some vector
$\vec{w}=(\vec{c_1}^{\perp},...,\vec{c_r}^{\perp})\in
(\mathcal{C}_1^{\perp}\times ...\times\mathcal{C}_r^{\perp})$.
Thus for every
$\vec{c}=(\vec{c_1},...,\vec{c_r})\widehat{C}\in\mathcal{C}_T$, we
see that
$$\vec{v}\cdot\vec{c}=\vec{w}C_{t}^{-1}\vec{c}_t=(\vec{c_1}^{\perp},...,\vec{c_r}^{\perp})\widehat{C}_{t}^{-1}
((\vec{c_1},...,\vec{c_r})\widehat{C})_t=$$
$$=(\vec{c_1}^{\perp},...,\vec{c_r}^{\perp})(\vec{c_1},...,\vec{c_r})_t=
\vec{c_1}^{\perp}\cdot\vec{c_1}+...+ \vec{c_r}^{\perp}\cdot\vec{c_r}=0,$$
i.e. $\mathcal{C}\subseteq {\mathcal{C}_T}^{\perp}$. Since $\dim \mathcal{C} =
\dim {\mathcal{C}_T}^{\perp}$, we conclude $\mathcal{C}={\mathcal{C}_T}^{\perp}$.

Finally, the second part of the statement follows easily from the
first one. \hfill $\square$

\begin{rem}\label{remark dual codes}
In the commutative case, the above result is useful to construct the Euclidean dual code of any product $T$-code and to
calculate its minimum Hamming distance (see Remark \ref{semilinear-rem} and Theorem \ref{rem Ore}). In fact, under the hypothesis that
either $\theta =id$, or $\theta\neq id$ and $R_n=R/Rf_1\times ...\times R/Rf_r$ with
$Rf_1,...,Rf_r$ pairwise coprime two-sided ideals of $R$, Theorem \ref{Euclidean dual semilinear} together with Remark \ref{remark commutative case},
Theorem \ref{rem Ore} and Proposition \ref{prop Ore} allow us to find the Euclidean dual code of a $T$-code.
\end{rem}

\noindent Finally, we obtain the following characterization of Euclidean dual codes of $T$-codes.

\begin{prop}\label{euclidean dual proposition}
Let $\mathcal{C}_T\subseteq\mathbb{F}_q^n$ be a $T$-code invariant under
a $\theta$-semi-linear transformation $T=\Theta\circ \overline{M}$.
Then the Euclidean dual code $\mathcal{C}_T^{\perp}$ is
a $T'$-code, where $T'=\Theta^{-1}\circ ({\overline{M}_t})_{\theta^{-1}}$.
\end{prop}

\noindent\textit{Proof}. If $\vec{a}\in\mathcal{C}_T^{\perp}$, then for every $\vec{c}\in\mathcal{C}_T$ we have
$$(\vec{a}\overline{M}_t)\cdot(\vec{c}\ \Theta)=\vec{a}(\vec{c}\ \Theta\circ \overline{M})_t=\vec{a}\cdot (\vec{c}\ T)=0.$$
Thus by Lemma \ref{lemma-euclidean} we deduce that
$$(\vec{a}\ T')\cdot\vec{c}=(\vec{a}\Theta^{-1}\circ ({\overline{M}_t})_{\theta^{-1}})\cdot\vec{c}=(\vec{a}\overline{M}_t\circ\Theta^{-1})\cdot\vec{c}=0,$$
for every $\vec{c}\in\mathcal{C}_T$, i.e. $\mathcal{C}_T^{\perp}$ is invariant under the semi-linear transformation $T'$. \hfill $\square$

\medskip

\subsection{Quasi-Euclidean duals}

In this subsection we introduce the new concept of quasi-Euclidean
dual codes and some of their properties related to the Euclidean dual
codes. Before to do this, we have to define a special injective
map for module $\theta$-codes.

\subsubsection{An injective map for an $f$-module $\theta$-code.}

Given a polynomial $f\in R$ of degree $n\geq 2$, we present here
an algorithm to show that there exists always a suitable integer
$m\geq n$ such that $X^m-1$ is a right multiple of $f$. This will
allow us to construct an immersion map of the code space
$\mathbb{F}_q^n$ into an $\mathbb{F}_q^m$ which will be useful for
the definition of quasi-Euclidean dual codes of a product $T$-code.

From now on, write
$$f=(-1)^n(X^n-\sum_{i=0}^{n-1}f_iX^i) \in R$$ and consider the
right division
$$X^n-1=f\cdot q_n + r_n,$$
where $q_n,r_n\in R$ and $0\leq\deg r_n<\deg f$. Assume that $r_n$
is not equal to zero, otherwise we are done.

Let $k$ be an integer such that $k> n$ and consider again the
right divisions
$$X^k-1=f\cdot q_k + r_k,$$
where $q_k,r_k\in R$ and $0\leq\deg r_k<\deg f=n$. Since there are
at most $q^{n+1}$ distinct polynomials $r_k$, we see that
for some $k_2> k_1\geq n$ we get $r_{k_1}=r_{k_2}$. Thus we obtain
that
$$X^{k_1}\cdot (X^{k_2-k_1}-1)=(X^{k_2-k_1}-1)\cdot X^{k_1}=f\cdot(q_{k_2}-q_{k_1}) .$$
Put $q':=q_{k_2}-q_{k_1}$ and note that $q'\neq 0\in R$. This
shows that $X=0$ is a root of $f\cdot q'$. Since $f(0)\neq 0$, we
deduce that $q'(0)=0$. Hence $q'=q_1\cdot X$ for some $q_1\in R$.

Thus we have
$$(X^{k_2-k_1}-1)\cdot X^{k_1}=f\cdot q_1\cdot X$$ and since $R$ has no zero divisors, we can deduce that
$$(X^{k_2-k_1}-1)\cdot X^{k_1-1}=f\cdot q_1$$ where $q_1\in R$. By an inductive argument, we can conclude that
$$X^{k_2-k_1}-1=f\cdot q'' $$ for some $q''\in R$. This shows that there exists always an integer $t\geq n$ such that
$X^t-1=f\cdot q_f$ for some non-zero $q_f=\sum_{i=0}^{m-n}q_iX^i\in
R$.

From now on, we denote by $m$ the minimum integer such that $m\geq
n$ and $X^m-1$ is a right multiple of $f$, i.e.

\begin{equation}
m:=\min\left\{i\in\mathbb{N}\ |\ X^i-1=f\cdot p\ \mathrm{for\
some}\ p\in R\right\}\ . \tag{**}
\end{equation}

\medskip

\noindent In this case, we write
$$X^m-1=f\cdot q_f\ .$$

\medskip

\noindent Moreover, by the above construction, we have
$$n\leq m\leq q^n+n-2.$$
Let us introduce the following ring isomorphism
$\Theta:\ R\to R$ defined as
$$(\sum_{i=0}^{t}a_iX^i)\Theta:=\sum_{i=0}^{t}\theta(a_i)X^i\ .$$

\begin{lem}\label{m}
Put
$$m^*:=\min\left\{j\in\mathbb{N}\ |\ X^j-1=p\cdot f^*\ \mathrm{for\ some}\ p\in R\right\}\ ,$$
where $f^*:=1-\sum_{i=1}^{n}\theta^{i}(f_{n-i})X^i \in R$. Then
$m^*=m$.
\end{lem}

\noindent\textit{Proof}.
Let $X^m-1=f\cdot q_f$. By \cite[Lemma 1(1)]{BU1} we know that
$X^m-1=(1-X^m)^*=(f\cdot (-q_f))^*=q'\cdot f^*$ for some $q'\in R$. This implies that $m\geq m^*$.
On the other hand, let $X^{m^*}-1=q_{f^*}\cdot f^*$. By
\cite[Lemma 1]{BU1} we see that
$$X^{m^*}-1=(1-X^{m^*})^*=((-q_{f^*})\cdot f^*)^*=$$
$$=((f^*)^*){\Theta}^{m^*-n}\cdot q''=((f){\Theta}^{n}){\Theta}^{m^*-n}\cdot q''=(f){\Theta}^{m^*}\cdot q''$$ for some $q''\in R$.
Hence we get
$$X^{m^*}-1=(X^{m^*}-1){\Theta}^{-m^*}=((f){\Theta}^{m^*}\cdot q''){\Theta}^{-m^*}=f\cdot (q''){\Theta}^{-m^*},$$
i.e. $m^*\geq m$. This gives $m^*=m$. \hfill $\square$

\bigskip

\noindent The following Magma \cite{M} program enables us to calculate the integer
$m$ as in $(**)$ for any polynomial $f\in\mathbb{F}_a[X;\theta] :$

\begin{verbatim}
F<w>:=GF(a); 
PeriodNC:=function(qq,g) 
 R<x>:=TwistedPolynomials(F:q:=qq); 
 f:=R!g; 
 n:=Degree(f)-1; 
 repeat n:=n+1; 
  _,r:=Quotrem(X^n-1,f); 
  until r eq R![0]; 
 return n; 
end function;
\end{verbatim}

\begin{rem}\label{m identity case}
If $\theta=id$, then the characteristic and minimal
polynomial of
$$A_c:=\left(
\begin{array}{c|ccc}
0 & 1 &  &  \\
\vdots &    & \ddots  &  \\
0 &   & & 1 \\
 \hline
f_{0} &  f_{1}  & \dots &  f_{n-1}
\end{array}\right)$$
are both equal to $f=(-1)^n(X^n-\sum_{i=0}^{n-1}f_iX^i) \in R$. Let
$m':=\min\left\{i\in\mathbb{N}\ |\ A_c^i=I\right\}$ and note that
the polynomial $X^{m'}-1$ is satisfied by $A_c$. Therefore, it
follows that there exists a polynomial
$q_{f}=\sum_{i=0}^{m-n}q_iX^i\in R$ such that $X^{m'}-1=f\cdot q_f
$. This gives $m=m'$, that is, $m=\min\left\{i\in\mathbb{N}\ |\
A_c^i=I\right\}.$ In this case, the following Magma~\cite{M} program gives us directly the integer $m=m'$ in $\mathbb{F}_a[X] :$
\begin{verbatim}
F<w>:=GF(a);
P<x>:=PolynomialRing(F);
PeriodC := function(f)
 return Order(CompanionMatrix(f));
end function;
\end{verbatim}
\end{rem}

\medskip

The following example shows that Remark \ref{m identity case} does
not hold in general when $\theta$ is not equal to the identity of
$\mathbb{F}_q$.

\begin{ex}\label{example1}
In $\mathbb{F}_{4}^3$, where
$\mathbb{F}_{4}=\mathbb{F}_{2}[\alpha]$ with
$\alpha^2+\alpha+1=0$, consider the polynomial $f=X^3+\alpha X+1$
associated to the matrix
$$\left(
\begin{array}{ccc}
0 & 1 & 0 \\
0 & 0 & 1 \\
1 & \alpha & 0
\end{array}\right)$$
and define $\theta(x)=x^2$ for any $x\in \mathbb{F}_{4}$.
It follows that $\min\left\{i\in\mathbb{N}\ |\
A_c^i=I\right\}=21$. Moreover, we have $X^{21}-1=f\cdot q+r,$
where
$$q=X^{18}+\alpha^2 X^{16}+X^{15}+\alpha X^{14}+X^{13}+X^{10}
+\alpha^2 X^{8}+X^7+\alpha X^6+X^5+X^2+\alpha^2$$ and
$r=X^{2}+\alpha^2X+\alpha\neq 0.$ This shows that
$m\neq\min\left\{i\in\mathbb{N}\ |\ A_c^i=I\right\}$.
Moreover, we get $m=8(<21)$. Hence $X^8-1=(X^3+\alpha X+1)\cdot
q_f$ with $q_f=X^5+\alpha^2X^3+X^2+\alpha X+1$.
\end{ex}

In connection with the above arguments, we have the following results.

\begin{prop}\label{lemma}
Let $m$ be an integer as in $(**)$ and let $P$ be the $m\times m$
matrix
$$\left(
\begin{array}{c|ccc}
0 & 1 &  &  \\
\vdots &    & \ddots  &  \\
0 &   & & 1 \\
 \hline
1 &  0  & \dots &  0
\end{array}\right) .$$
Denote by $\vec{q_f}:=(q_0,...,q_{m-n},0,...,0)\in\mathbb{F}_q^m$,
where the $q_i$'s are the coefficients of $q_f\in R$ as in $(**)$.
Then there exists a commutative diagram
\begin{displaymath}
    \xymatrix{
        \mathbb{F}_q^n \ar[r]^{i} \ar[d]_{\pi} & \mathbb{F}_q^m \ar[d]^{\pi'} \\
        R_n \ar[r]_{j}      & R_m }
\end{displaymath}
such that $\pi'\circ i=j\circ\pi$, where $R_n:=R/Rf$,
$R_m:=R/R(X^m-1)$, $i(\vec{v}):=\vec{v}Q$ with $Q$ the matrix
$$\left(
\begin{array}{c}
\vec{q_f} \\
(\vec{q_f})(\Theta\circ P) \\
(\vec{q_f})(\Theta\circ P)^2 \\
... \\
(\vec{q_f})(\Theta\circ P)^{n-1}
\end{array}\right)$$
and $j(a+Rf):=(a\cdot q_{f})+R(X^m-1)$ for any $a\in R$.
\end{prop}

\noindent\textit{Proof}. By using the canonical basis of
$\mathbb{F}_q^n$, the statement follows easily from the linearity of the
maps $i,j,\pi$ and $\pi'$. \hfill $\square$

\begin{prop}\label{prop}
With the same notation as in {\em Proposition \ref{lemma}}, for any
$\vec{c}\in\mathbb{F}_q^n$ and $k\in\mathbb{N}$ we have
$$i((\vec{c})(\Theta\circ A_c)^k)=(i(\vec{c}))(\Theta\circ P)^k,$$
where $A_c$ is the matrix defined in {\em Remark \ref{m identity case}}.
\end{prop}

\noindent\textit{Proof}. Let $\vec{c}\in\mathbb{F}_q^n$. By Proposition
\ref{lemma}, we have the following two commutative diagrams:
\begin{displaymath}
    \xymatrix{
        \vec{c} \ar[r]^{i} \ar[d]_{\pi} & i(\vec{c}) \ar[d]^{\pi'} \\
        \pi(\vec{c}) \ar[r]_{j}      & j(\pi(\vec{c})) }
\end{displaymath}
where $j(\pi(\vec{c}))=\pi'(i(\vec{c}))$, and
\begin{displaymath}
    \xymatrix{
       (\vec{c})(\Theta\circ A_c)^k \ar[r]^{i} \ar[d]_{\pi} & i((\vec{c})(\Theta\circ A_c)^k) \ar[d]^{\pi'} \\
        X^k\cdot\pi(\vec{c}) \ar[r]_{j}      & j(X^k\cdot\pi(\vec{c})) }
\end{displaymath}
where $j(X^k\cdot\pi(\vec{c}))=\pi'(i((\vec{c})(\Theta\circ
A_c)^k))$. Since $\pi'$ is an isomorphism, by the commutative
diagram of Proposition \ref{lemma}, we obtain
$$i((\vec{c})(\Theta\circ A_c)^k)=(\pi')^{-1}(j(X^k\cdot\pi(\vec{c})))=(\pi')^{-1}(X^k\cdot\pi(\vec{c})\cdot q_f)=$$
$$=(\pi')^{-1}(X^k\cdot j(\pi(\vec{c})))=(\pi')^{-1}(X^k\cdot
\pi'(i(\vec{c})))=(\pi')^{-1}\circ\pi'((i(\vec{c}))(\Theta\circ
P)^k),$$ that is, $i((\vec{c})(\Theta\circ
A_c)^k)=(i(\vec{c}))(\Theta\circ P)^k$ for any $k\in\mathbb{N}$.
\hfill $\square$

\begin{rem}
The maps $i$ and $j$ in Proposition \ref{lemma} are injective. Moreover,
Proposition \ref{prop} shows that the image via $i$ of an
$f$-module $\theta$-code in $\mathbb{F}_q^n$ is a module
$\theta$-cyclic code in $\mathbb{F}_q^m$, where $m$ is defined as
in $(**)$ (or as in Remark \ref{m identity case}).
\end{rem}

Let $s$ be the order of $\theta$. From the above results, we can
deduce the following two consequences.

\begin{cor}\label{cor q}
Let $m$ be as in $(**)$. If $m=as+r$, $0\leq r<s$, then
$(\vec{q_f})\Theta^r=\vec{q_f}$.
\end{cor}

\noindent\textit{Proof}. Since $X^m-1=f\cdot q_f$ and $f,q_f$ are
monic polynomials, by \cite[Lemma 2(2)]{BU1} we see that
$X^m=1+(q_f)\Theta^m\cdot f$. Since $(\Theta\circ P)^m=
\Theta^m\circ P^m=\Theta^m$ and $\Theta^s$ is the identity, from
the following commutative diagram
\begin{displaymath}
    \xymatrix{
    \vec{e}_1(\Theta\circ A_c)^m  \ar[r]^{i} \ar[d]_{\pi} & i(\vec{e}_1)(\Theta\circ P)^m \ar[d]^{\pi'} \\
       X^m=1 \ar[r]_{j}      & j(1)=q_f  }
\end{displaymath}
we conclude that
$\vec{q_f}=(\pi')^{-1}(q_f)=i(\vec{e}_1)(\Theta\circ
P)^m=(\vec{q_f})\Theta^m=(\vec{q_f})\Theta^r.$ \hfill $\square$

\begin{cor}\label{cor m}
Let $f=(-1)^n(X^n-\sum_{i=0}^{n-1}f_iX^i) \in R$. If
$$(f_0,f_1,...,f_{n-1})\Theta^t\neq (f_0,f_1,...,f_{n-1})$$ for every integer $t$
such that $0<t<s$, then the order $s$ of $\Theta$ divides $m$.
\end{cor}

\noindent\textit{Proof}. Since $X^m-1=f\cdot q_f$, from Corollary
\ref{cor q} it follows that
$$f\cdot q_f=X^m-1=(X^m-1)\Theta^m=(f)\Theta^m\cdot (q_f)\Theta^m =(f)\Theta^m \cdot q_f,$$
i.e. $f=(f)\Theta^m$. Let $m=as+r$ with $0\leq r<s$.

Assume now that $r\neq 0$. Then we get
$f=(f)\Theta^m=(f)\Theta^r$, that is,
$$(f_0,f_1,...,f_{n-1})\Theta^r = (f_0,f_1,...,f_{n-1})$$ for some $0<r<s$, but this is a contradiction. Thus $r=0$ and $s$ divides $m$.
\hfill $\square$

\medskip

\begin{ex}\label{example2}
In $\mathbb{F}_{4}^5$, where
$\mathbb{F}_{4}=\mathbb{F}_{2}[\alpha]$ with $\alpha^2+\alpha+1=0$
and $\theta$ is the Frobenius map, consider the following two
polynomials:
\begin{enumerate}
\item[$(1)$] $f=X^5+X^3+X^2+1$; \quad $(2)\ g=X^5+X^2+1$.
\end{enumerate}
Note that in both cases the hypothesis of Corollary \ref{cor m} is
not satisfied. Moreover, we have $m=12$ in case $(1)$ and $m=31$ in case $(2)$.
\end{ex}

Finally, let us give here also some results about the integer $m$ in $(**)$ when $\theta$ is the identity of
$\mathbb{F}_q$.

\begin{rem}
Let $\mathbb{F}_q\subseteq\mathbb{K}$ be a finite extension of
$\mathbb{F}_q$ such that $f=\prod_{i=1}^{n}(X-a_i)$ with
$a_i\in\mathbb{K}$ and $A_c$ is diagonalizable over $\mathbb{K}$. If $m_i:=\min\left\{h_i\ |\ a_i^{h_i}=1
\right\}$, then $m=lcm(m_1,...,m_n)$.
\end{rem}

\begin{rem}
Let $p:=\mathrm{Char}(\mathbb{F}_q)$. If the polynomial $f$ has a
root of multiplicity $\geq 2$, then $X^m-1$ has a root of
multiplicity $\geq 2$. This shows that $gcd(m,p)\neq 1$ and since
$p$ is a prime number, we get $m\equiv 0\mod p$.
\end{rem}

The next two results give a more simple computation of $m$.

\begin{prop}
Denote by $\vec{f}:=(f_0,...,f_{n-1})$ and let
$$k:=\min\left\{h\in\mathbb{N}\cup\{0\} \ |\ \vec{f}A_c^h=\vec{e}_1\right\}.$$
Then $m=n+k$. In particular, we have $\deg q_f=k$.
\end{prop}

\noindent\textit{Proof}. For any $h=1,...,n$, we have
$$\vec{e}_h\ A_c^{n+k} = ((\vec{e}_hA_c^{n-h+1})A_c^k)A_c^{h-1}=((\vec{e}_nA_c)A_c^k)A_c^{h-1}=$$
$$=(\vec{f}A_c^k)A_c^{h-1}=\vec{e}_1A_c^{h-1}=\vec{e}_h.$$
Hence $A_c^{n+k}=I$ and for the minimality of $m$ we deduce that
$m\leq n+k$. Furthermore, since $A_c^m=I$ we get
$\vec{e}_1=((\vec{e}_1A_c^{n-1})A_c)A_c^{m-n}=(\vec{e}_nA_c)A_c^{m-n}=\vec{f}A_c^{m-n},$
that is, $\vec{f}A_c^{m-n}=\vec{e}_1$. So, by definition of $k$ we
can conclude that $k\leq m-n$, i.e. $m\geq n+k$. Finally, observe that $\deg q_f=m-n:=k$. \hfill $\square$

\bigskip

Let $p_0$ be the order of $\det A_c$. Since $A_c^m=I$, it follows
that $(\det A_c)^m=1$, i.e. $m\equiv 0\mod p_0$ with $p_0$ the
order of $\det A_c$. Denote by $B:=A_c^{p_0}$. From this it
follows immediately also the following

\begin{prop}
Let $m'$ be the minimum integer such
that $B^{m'}$ is the identity matrix.
Then $m=p_0m'$. In particular,
we have $\deg q_f=p_0m'-n$.
\end{prop}

When $\theta = id$, all the above results give the following

\smallskip

\noindent\textbf{Algorithm 1:}

\textbf{Input}: $f$
\begin{itemize}
\item Define $a_0:=\det A_c$; \item Compute the order $p_0$ of
$a_0$; \item Define $B:=A_c^{p_0}$; \item Find the rational
canonical form $B'$ of $B$; \item For any diagonal block $B_i$,
$i=1,...,s$, of $B'$ compute $m'_i=\min\left\{h\ |\ B_i^h=I
\right\}.$
\end{itemize}

\textbf{Output}: $m=lcm(m'_1,...,m'_s)\cdot p_0$.

\medskip

\noindent The following Magma \cite{M} program is an application of Algorithm 1 in $\mathbb{F}_a[X] :$

\begin{verbatim}
F<w>:=GF(a);
P<x>:=PolynomialRing(F);
Period := function(f) 
 d:=Degree(f); 
 A:=CompanionMatrix(f); 
 p:=Order(Determinant(A)); 
 _,_,E:=PrimaryRationalForm(A^p);
 ll:=[];
  for j in [1..#E] do 
   ll := ll cat [Order(CompanionMatrix(E[j][1]))];
  end for;
 return LCM(ll); 
end function;
\end{verbatim}

\subsubsection{Definition and basic properties of quasi-Euclidean dual codes}

Under the hypothesis $(*)$, write
$\mathbb{F}_q^n=\mathbb{F}_q^{n_1}\times ...
\times\mathbb{F}_q^{n_r}$ with $r\geq 1$ and $n=\sum_{k=1}^{r}n_k$.
From Proposition \ref{lemma}, we know that for every $k=1,...,r$ there exists a commutative
diagram
\begin{displaymath}
    \xymatrix{
        \mathbb{F}_q^{n_k} \ar[r]^{i_k} \ar[d]_{\pi_k} &  \mathbb{F}_q^{m_k} \ar[d]^{\pi_k'} \\
            R/Rf_k  \ar[r]_{j_k}   & R/R(X^{m_k}-1) }\ .
    \end{displaymath}
Consider the further commutative diagram:
\begin{displaymath}
    \xymatrix{
        \mathbb{F}_q^n \ar[dr]^{\overline{i}} \ar[d]_{\varphi :=C^{-1}} &  \\
        \mathbb{F}_q^{n}=\mathbb{F}_q^{n_1}\times ...\times\mathbb{F}_q^{n_r}  \ar[r]_{i}\ar[d]_{\pi}
            & \mathbb{F}_q^{m_1}\times ...\times\mathbb{F}_q^{m_r}=\mathbb{F}_q^{m} \ar[d]_{\pi'} \\
       R_n \ar[r]_{j} & R_m }
\end{displaymath}
where $n=\sum_{i=1}^{r}n_i$, $m=\sum_{i=1}^{r}m_i$ with the $m_i$'s as in $(*)$,
$f_i=(-1)^{n_i}(X^{n_i}-\sum_{j=0}^{n_i-1}f_{i,j}X^j) \in R$,
$$R_n:=R/Rf_1\times ... \times R/Rf_r,$$
$$R_m:=R/R(X^{m_1}-1)\times ... \times R/R(X^{m_r}-1),$$ $i(\vec{v}):=
\vec{v}\widehat{Q}$ with
$$\widehat{Q}:=\left(\begin{array}{cccc}
Q_1 &  &  &  \\ [2pt]
 & Q_2 &  &  \\  [2pt]
 &  & \ddots &  \\  [2pt]
 &  &  & Q_r \\  [2pt]
\end{array}\right)$$
and all the $Q_i$'s are matrices $n_i\times m_i$ as in Proposition
\ref{lemma}, $\pi=(\pi_1,...,\pi_r)$ with
$\pi_i:\mathbb{F}_q^{n_i}\to R/Rf_i$, $\pi'=(\pi_1',...,\pi_r')$
with $\pi_i':\mathbb{F}_q^{m_i}\to R/R(X^{m_i}-1)$ and
$$j(p_1,...,p_r):=(p_1\cdot q_{f_1},...,p_r\cdot q_{f_r})$$ with all the $q_{f_i}$'s polynomials in $R$ as in Proposition \ref{lemma}.

Denote by $\mathcal{I}$ the image of $\overline{i}=i\circ\varphi$
and define
$$B:=C^{-1}\widehat{Q}\ {\widehat{Q}}_t(C^{-1})_t,$$
where $M_t$ is the transpose of a matrix $M$. Note that $B$ is a
symmetric matrix.

Let $r:=\mathrm{rk} B$ be the rank of $B$ and observe that
$$r=\mathrm{rk}(\widehat{Q}\cdot {\widehat{Q}}_t )=n-\dim (\mathrm{Ker\ }{\widehat{Q}}_t\cap\mathcal{I})$$ with $0\leq r\leq n$.

\begin{defn}\label{definition}
Let $T$ be a semi-linear transformation of $\mathbb{F}_q^n$ as in $(*)$. We define
the quasi-Euclidean scalar product $\cdot_{*}$ on
$\mathbb{F}_q^n$ as
$\vec{a}\cdot_{*}\vec{b}:=\vec{a}B\vec{b}_t$ for any
$\vec{a},\vec{b}\in\mathbb{F}_q^n$, and we denote by $\mathcal{C}^{*}$ the
linear \textit{quasi-Euclidean dual code} of a linear code $\mathcal{C}$ with respect to
$\cdot_{*}$, i.e.
$$\mathcal{C}^{*}:=\left\{\vec{x}\in\mathbb{F}_q^n\ |\ \vec{x}\cdot_{*}\vec{c}=0\ \mathrm{for\ every}\ \vec{c}\in \mathcal{C}\right\}.$$
\end{defn}

\begin{thm}\label{prop quasi euclidean}
Let $\mathcal{C}\subset\mathbb{F}_q^n$ be a linear code. Then we
have the following properties:
\begin{enumerate}
\item[(i)] $\mathcal{C}^{*}=(\mathcal{C}\cdot B)^{\perp}$;
\item[(ii)] $\dim \mathcal{C}^{*}=\dim
\mathcal{C}^{\perp}+\dim(\mathcal{C}\cap\mathrm{Ker\ }B)$;
\item[(iii)] $\mathcal{C}^{*} B=\mathcal{C}^{\perp}\cap (\mathrm{Im}\ B)$, where $\mathrm{Im}\ B:=\{\vec{y}\in\mathbb{F}_q^n\ |\ \vec{y}=\vec{x}B\ \mathrm{for\ some\ } \vec{x}\in\mathbb{F}_q^n \}$;
\item[(iv)] $(\mathcal{C}^{*})^{*}=\mathcal{C}+\mathrm{Ker\ }B$, where $\mathrm{Ker\ }B:=\{\vec{x}\in\mathbb{F}_q^n\ |\ \vec{x}B=\vec{0} \}$;
\item[(v)]
$\overline{i}(\mathcal{C}^{*})=\overline{i}(\mathcal{C})^{\perp}\cap\mathcal{I}
=\overline{i}(\mathcal{C}+\mathrm{Ker\
}B)^{\perp}\cap\mathcal{I}$; \item[(vi)]
$(\mathbb{F}_q^n)^*=\mathrm{Ker\ }B=(\mathrm{Im}\ B)^{\perp},\
(\mathrm{Ker\ }B)^*=\mathbb{F}_q^n,\ (\mathrm{Ker\
}B)^{**}=\mathrm{Ker\ }B$.
\end{enumerate}
\end{thm}

\noindent\textit{Proof}. (i) To prove
$\mathcal{C}^*=(\mathcal{C}\cdot B)^{\perp}$, we observe that
\begin{equation*}
\begin{split}
\vec{w}\in (\mathcal{C}\cdot B)^{\perp} & \iff \vec{w}\cdot (\vec{c}B)=0, \quad \forall \vec{c}\in \mathcal{C} \\
 & \iff \vec{w}B_t\vec{c}_t=0, \quad \forall \vec{c}\in \mathcal{C} \\
 & \iff \vec{w}B\vec{c}_t=0, \quad \forall \vec{c}\in \mathcal{C} \\
 & \iff \vec{w}\cdot_*\vec{c}=0, \quad \forall \vec{c}\in \mathcal{C} \\
 & \iff \vec{w}\in \mathcal{C}^*\ .
\end{split}
\end{equation*}

(ii) This follows easily from $$\dim (\mathcal{C}\cdot B)=\dim
\mathcal{C} - \dim (\mathcal{C}\cap\mathrm{Ker\ }B)$$ and $\dim
\mathcal{C}^* = n -\dim (\mathcal{C}\cdot B)$.

\medskip

(iii) If $\vec{x}\in \mathcal{C}^*B$, then $\vec{x}\in\mathrm{Im}\ B$ and
$\vec{x}=\vec{c^*}B$ for some $\vec{c^*}\in \mathcal{C}^*$. Hence
for every $\vec{c}\in \mathcal{C}$ we get
$$\vec{x}\cdot \vec{c}=\vec{c^*}B\cdot \vec{c}=\vec{c^*}\cdot_* \vec{c}=0,$$ i.e. $\mathcal{C}^*B\subseteq \mathcal{C}^{\perp}\cap (\mathrm{Im}\ B)$.
On the other hand, let $\vec{y}\in \mathcal{C}^{\perp}\cap (\mathrm{Im}\
B)$. Then $\vec{y}=\vec{v}B\in \mathcal{C}^{\perp}$ for some
$\vec{v}\in\mathbb{F}_q^n$. Thus for any $\vec{c}\in \mathcal{C}$
we have
$$\vec{v}\cdot_*\vec{c}=\vec{v}B\vec{c}_t=\vec{y}\cdot \vec{c}=0,$$ that is, $\mathcal{C}^{\perp}\cap (\mathrm{Im}\ B)\subseteq \mathcal{C}^*B$.

\medskip

(iv) Let $\vec{x}=\vec{c}+\vec{b}\in \mathcal{C}+\mathrm{Ker\ }B$.
Then for every $\vec{c^*}\in \mathcal{C}^*$ by $(i)$ we have
$$\vec{x}\cdot_*\vec{c^*}=\vec{x}B\vec{c^*}_t=(\vec{c}B+\vec{b}B)\cdot \vec{c^*}=(\vec{c}B)\cdot \vec{c^*}=0,$$
i.e. $\mathcal{C}+\mathrm{Ker\ }B\subseteq (\mathcal{C}^*)^*$.
Let $\vec{v}\in (\mathcal{C}^*)^*$. Then for any
$\vec{x}\in\mathcal{C}^*$ we get
$$\vec{v}B\cdot\vec{x}=\vec{v}B\vec{x}_t=\vec{v}\cdot_*\vec{x}=0,$$
i.e. $\vec{v}B\in (\mathcal{C}^*)^{\perp}=\mathcal{C}B$. Thus
there exists a $\vec{c}\in\mathcal{C}$ such that
$\vec{v}B=\vec{c}B$. This implies that
$(\vec{v}-\vec{c})B=\vec{0}$, that is,
$\vec{v}-\vec{c}\in\mathrm{Ker\ }B$ and $\vec{v}=\vec{c}+\vec{b}$
for some $\vec{b}\in\mathrm{Ker\ }B$.

\medskip

(v) If $\vec{x}\in \overline{i}(\mathcal{C}^*)$, then
$\vec{x}=\overline{i}(\vec{v})=\vec{v}C^{-1}\widehat{Q}\in\mathcal{I}$
for some $\vec{v}\in \mathcal{C}^*$. Hence for every
$\vec{c}\in\mathcal{C}$ and $\vec{b}\in\mathrm{Ker\ }B$, we have
$$\vec{x}\cdot \overline{i}(\vec{c}+\vec{b})=\vec{x}\cdot \overline{i}(\vec{c})
+\vec{x}\cdot
\overline{i}(\vec{b})=\vec{v}\cdot_*\vec{c}+\vec{v}\cdot
(\vec{b}B)=0,$$ that is, $\overline{i}(\mathcal{C}^*)\subseteq
\overline{i}(\mathcal{C}+\mathrm{Ker\ }B)^{\perp}\cap\mathcal{I}$.
Now, let $\vec{x}\in \overline{i}(\mathcal{C}+\mathrm{Ker\
}B)^{\perp}\cap\mathcal{I}$, i.e.
$\vec{x}=\overline{i}(\vec{v})=\vec{v}C^{-1}\widehat{Q}\in
\overline{i}(\mathcal{C}+\mathrm{Ker\ }B)^{\perp}\subseteq
\overline{i}(\mathcal{C})^{\perp}$ for some
$\vec{v}\in\mathbb{F}_q^n$. Thus for every $\vec{y}\in
\mathcal{C}$ we have
$$\vec{v}\cdot_* \vec{y}=\vec{v}B\vec{y}_t=(\vec{v}C^{-1}\widehat{Q})
(\vec{y}C^{-1}\widehat{Q})_t =\vec{x}\cdot
\overline{i}(\vec{y})=0,$$ i.e. $\vec{v}\in \mathcal{C}^*$. Hence
we get $\vec{x}=\overline{i}(\vec{v})\in
\overline{i}(\mathcal{C}^*)$, that is,
$\overline{i}(\mathcal{C}+\mathrm{Ker\
}B)^{\perp}\cap\mathcal{I}\subseteq \overline{i}(\mathcal{C}^*)$.

Let us prove now that $\overline{i}(\mathcal{C}^{*})$ is also
equal to $\overline{i}(\mathcal{C})^{\perp}\cap\mathcal{I}$. Let
$\vec{x}\in\overline{i}(\mathcal{C}^{*})$. Then
$\vec{x}=\overline{i}(\vec{c^{*}})\in\mathcal{I}$ for some vector
$\vec{c^{*}}\in\mathcal{C}^{*}$. Therefore for every
$\vec{c}\in\mathcal{C}$ we have
$$\vec{x}\cdot\overline{i}(\vec{c})=
\overline{i}(\vec{c^{*}})\cdot \overline{i}(\vec{c})=
\vec{c^{*}}\cdot_* \vec{c}=0,$$ i.e.
$\vec{x}\in\overline{i}(\mathcal{C})^{\perp}\cap\mathcal{I}$. On
the other hand, let
$\vec{y}\in\overline{i}(\mathcal{C})^{\perp}\cap\mathcal{I}$. Then
$\vec{y}=\overline{i}(\vec{z})\in\mathcal{I}$ for some
$\vec{z}\in\mathbb{F}_q^n$ and for every $\vec{c}\in\mathcal{C}$
we get
$$0=\overline{i}(c)\cdot\vec{y}=\overline{i}(c)\cdot\overline{i}(z)=
\vec{c}\cdot_*\vec{z}.$$ Hence $\vec{z}\in\mathcal{C}^*$, i.e.
$\vec{y}\in\overline{i}(\mathcal{C}^*)$.

\medskip

(vi) Since $(\{\vec{0}\})^*=\mathbb{F}_q^n$, the equalities
$(\mathbb{F}_q^n)^*=(\mathrm{Im}\ B)^{\perp}$ and
$(\mathbb{F}_q^n)^*=\mathrm{Ker\ }B$ follow easily from (i) with
$\mathcal{C}=\mathbb{F}_q^n$ and from (iv) with
$\mathcal{C}=\{\vec{0}\}$ respectively. Finally, by taking
$\mathcal{C}=\mathrm{Ker\ }B$, the equalities $(\mathrm{Ker\
}B)^*=\mathbb{F}_q^n$ and $(\mathrm{Ker\ }B)^{**}=\mathrm{Ker\ }B$
are immediate consequences of (i) and (iv), respectively. \hfill
$\square$

\begin{cor}\label{cor r=n}
Let $\mathcal{C}\subseteq \mathbb{F}_q^n$ be a linear code. If $r=n$,
then we have
\begin{enumerate}
\item[(j)] $\mathcal{C}^{*}=\mathcal{C}^{\perp}\cdot B^{-1}$;
\item[(jj)] $\dim \mathcal{C}^{*}=\dim \mathcal{C}^{\perp}$;
\item[(jjj)] $(\mathcal{C}^{*})^{*}=\mathcal{C}$; \item[(jv)]
$\overline{i}(\mathcal{C}^{*})=\overline{i}(\mathcal{C})^{\perp}
\cap\mathcal{I}$; \item[(v)] $(\mathbb{F}_q^n)^*=\{\vec{0}\},\
(\{\vec{0}\})^*=\mathbb{F}_q^n$.
\end{enumerate}
\end{cor}

\begin{rem}\label{rem r=n,r=0}
When $r=n$, by Corollary \ref{cor r=n} (j) we can easily obtain a generator matrix of
$\mathcal{C}^*$ by multiplying the parity check matrix
of $\mathcal{C}$ with the matrix $B^{-1}$. Moreover, when $r=0$, we see that $B$ is the null matrix and in this case $\widehat{Q}$ represents
a generator matrix of an euclidean self-orthogonal code $\mathcal{C}$ (i.e. $\mathcal{C}\subseteq\mathcal{C}^{\perp}$) of dimension $n$ in $\mathbb{F}_q^m$.
\end{rem}

\begin{rem}
From Theorem \ref{prop quasi euclidean} (vi), it follows that $\mathrm{Ker\ }B\subseteq \{\vec{v}\}^*$ for any $\vec{v}\in\mathbb{F}_q^n$.
In particular, we deduce that $\mathrm{Ker\ }B\subseteq\mathcal{C}^*$ for any linear code $\mathcal{C}\subseteq\mathbb{F}_q^n$.
\end{rem}

\begin{ex}\label{example3}
In $\mathbb{F}_{4}^3$, where $\mathbb{F}_{4}=\mathbb{F}_{2}[\alpha]$ with $\alpha^2+\alpha+1=0$, consider the following four polynomials:
\begin{enumerate}
\item[$(1)$] $f_0=X^3+X^2+1$; \quad $(2)\ f_1=X^3+\alpha^2 X^2+\alpha^2 X+\alpha$;
\item[$(3)$] $f_2=X^3+X^2+\alpha X+\alpha^2$; \quad $(4)\ f_3=X^3+\alpha^2$.
\end{enumerate}
Note that $m=7$ for the first case, while $m=6$ for the other cases. Then
$$X^6-1=f_1\cdot q_{f_1}=f_2\cdot q_{f_2}=f_3\cdot q_{f_3},\ X^7-1=f_0\cdot q_{f_0}$$
where
\begin{equation*}
\begin{split}
q_{f_0} =X^4+X^3 + X^2 + 1, & \quad q_{f_1} = X^3 + \alpha X^2 + \alpha^2 X + \alpha^2, \\
q_{f_2} =X^3 + X^2 + \alpha X + \alpha, & \quad q_{f_3} =X^3 + \alpha .
\end{split}
\end{equation*}
Therefore this gives
$$Q_0=\left(
\begin{array}{ccccccc}
1 & 0 & 1 & 1 & 1 & 0 & 0 \\
0 & 1 & 0 & 1 & 1 & 1 & 0 \\
0 & 0 & 1 & 0 & 1 & 1 & 1
\end{array}\right), \quad
Q_1=\left(
\begin{array}{cccccc}
\alpha^2 & \alpha^2 & \alpha & 1 & 0 & 0 \\
0 & \alpha & \alpha & \alpha^2 & 1 & 0 \\
0 & 0 & \alpha^2 & \alpha^2 & \alpha & 1
\end{array}\right),$$
$$Q_2=\left(
\begin{array}{cccccc}
\alpha & \alpha & 1 & 1 & 0 & 0 \\
0 & \alpha^2 & \alpha^2 & 1 & 1 & 0 \\
0 & 0 & \alpha & \alpha & 1 & 1
\end{array}\right), \quad
Q_3=\left(
\begin{array}{cccccc}
\alpha & 0 & 0 & 1 & 0 & 0 \\
0 & \alpha^2 & 0 & 0 & 1 & 0 \\
0 & 0 & \alpha & 0 & 0 & 1
\end{array}\right),$$
and
$$B_0=\left(
\begin{array}{ccc}
0 & 0 & 0 \\
0 & 0 & 0 \\
0 & 0 & 0
\end{array}\right),\
B_1=\left(
\begin{array}{ccc}
\alpha & 1 & \alpha \\
1 & \alpha^2 & 1 \\
\alpha & 1 & \alpha
\end{array}\right),$$
$$B_2=\left(
\begin{array}{ccc}
0 & \alpha^2 & 0 \\
\alpha^2 & 0 & \alpha \\
0 & \alpha & 0
\end{array}\right),\
B_3=\left(
\begin{array}{ccc}
\alpha & 0 & 0 \\
0 & \alpha^2 & 0 \\
0 & 0 & \alpha
\end{array}\right),
$$
with $\mathrm{rk\ }B_i=i$ for $i=0,...,3$.
Observe that
from Remark \ref{rem r=n,r=0} it follows that $Q_0$ is the generator matrix of an euclidean self-orthogonal code
(in fact, an euclidean self-orthogonal cyclic code) of type $[7,3]_4$ with minimum Hamming distance equal to three.
\end{ex}

\begin{cor}
Let $\mathcal{C}$ be a linear code in $\mathbb{F}_q^n$. Then
$$\mathcal{C}\subseteq \mathcal{C}^*\iff \overline{i}(\mathcal{C})\subseteq \overline{i}(\mathcal{C})^{\perp},$$ i.e.
$\mathcal{C}$ is self-ortogonal with respect to $\cdot_*$ if and
only if $\overline{i}(\mathcal{C})$ is self-ortogonal with respect
to $\cdot$.
\end{cor}

\noindent\textit{Proof}. Since $\overline{i}$ is injective, the
statement is an immediate consequence of Theorem \ref{prop
quasi euclidean} (v) and the following equivalence: \
$\overline{i}(\mathcal{C})\subseteq \overline{i}(\mathcal{C})^{\perp}\cap
\mathcal{I}\iff \overline{i}(\mathcal{C})\subseteq
\overline{i}(\mathcal{C})^{\perp}.$ \hfill $\square$

\medskip

\begin{lem}\label{lem semilinear}
For any $\vec{c}\in\mathbb{F}_q^n$ and $k\in\mathbb{N}$, we have
$$\overline{i}(\vec{c}\ T^k)=\overline{i}(\vec{c})(\Theta\circ \widehat{P})^k,$$
where
$$\widehat{P}:=\left(\begin{array}{cccc}
P_1 &  &  &  \\ [2pt]
 & P_2 &  &  \\  [2pt]
 &  & \ddots &  \\  [2pt]
 &  &  & P_r \\  [2pt]
\end{array}\right)$$
and the $P_i$'s are the $m_i\times m_i$ matrices as in {\em Proposition \ref{lemma}} for every $i=1,...,r$.
\end{lem}

\noindent\textit{Proof}. It is sufficient to prove the statement
for $k=1$. Thus, for every $\vec{c}\in\mathbb{F}_q^n$, let
$\vec{v}=(\vec{v}_1,...,\vec{v}_r)\in\mathbb{F}_q^{n_1}\times ...
\times \mathbb{F}_q^{n_r}$ be the vector such that
$\vec{c}=\vec{v}C$. By definition and Proposition \ref{prop} we
have
$$\overline{i}(\vec{c}\ T)=i(\vec{c}\ T C^{-1})= i(\vec{c}\ C^{-1}(\Theta\circ D) CC^{-1})= i((\vec{v})(\Theta\circ D))=$$
$$=i((\vec{v}_1\Theta\circ M_1,...,\vec{v}_r\Theta\circ M_r))= ((\vec{v}_1)(\Theta\circ M_1)Q_1,...,(\vec{v}_r)(\Theta\circ M_r)Q_r)= $$
$$=(i_1(\vec{v}_1\Theta\circ M_1),...,i_r(\vec{v}_r\Theta\circ M_r))=(i_1(\vec{v}_1)(\Theta\circ P_1),...,i_r(\vec{v}_r)(\Theta\circ P_r))= $$
$$=(i_1(\vec{v}_1),...,i_r(\vec{v}_r))(\Theta\circ \widehat{P})=(\vec{v}_1Q_1,...,\vec{v}_rQ_r)(\Theta\circ \widehat{P})=$$
$$=(\vec{v}_1,...,\vec{v}_r)\widehat{Q}(\Theta\circ \widehat{P})=(\vec{v})\widehat{Q}(\Theta\circ \widehat{P})=i(\vec{v})(\Theta\circ \widehat{P}),$$
that is, $\overline{i}(\vec{c}\ T)=i(\vec{v})(\Theta\circ
\widehat{P})=\overline{i}(\vec{c})(\Theta\circ \widehat{P}).$
\hfill $\square$

\begin{cor}\label{T-cor 0}
Let $\mathcal{C}\subseteq\mathbb{F}_q^n$ be a linear code. Then

\smallskip

$\mathcal{C}$ is a $T$-code $\iff$ $\overline{i}(\mathcal{C})$ is a linear code invariant under $\Theta\circ \widehat{P}$.
\end{cor}

\noindent\textit{Proof}. From Lemma \ref{lem semilinear} it follows that

\smallskip

$\mathcal{C}$ is a product $T$-code $\iff$ $\varphi (\mathcal{C})$
is a linear code invariant by $\Theta\circ D$ $\iff$ $i(\varphi
(\mathcal{C}))=\overline{i}(\mathcal{C})$ is a linear code invariant by $\Theta\circ
\widehat{P}$. \hfill $\square$

\begin{cor}\label{T-cor}
Let $\mathcal{C}=(\mathcal{C}_1\times
...\times\mathcal{C}_r)\widehat{C}C$ be a linear code in
$\mathbb{F}_q^n=\mathbb{F}_q^{n_1}\times ...\times\mathbb{F}_q^{n_r}$, where $\widehat{C}$ is an invertible matrix and
$\mathcal{C}_i\subseteq\mathbb{F}_q^{n_i}$ is a linear code for every $i=1,...,r$. If there exists an invertible matrix $\overline{C}$ such that
$\widehat{C}(\widehat{Q}\widehat{Q}_t)=(\widehat{Q}\widehat{Q}_t)\overline{C}$, then $\mathcal{C}^{*}=(\mathcal{C}_1^*\times
...\times\mathcal{C}_r^*)\overline{C}_t^{-1}C$, where $\mathcal{C}_i^*\subseteq\mathbb{F}_q^{n_i}$ is the quasi-Euclidean dual code of $\mathcal{C}_i$
for every $i=1,...,r$. In particular, if $\mathcal{C}=(\mathcal{C}_1\times
...\times\mathcal{C}_r)C$ is a product $T$-code, then $\mathcal{C}^{*}=(\mathcal{C}_1^*\times
...\times\mathcal{C}_r^*)C$.
\end{cor}

\noindent\textit{Proof}. By Theorem \ref{prop quasi
euclidean}(i) and Theorem \ref{Euclidean dual semilinear}, we have
$$\mathcal{C}^{*}=(\mathcal{C}B)^{\perp}=((\mathcal{C}_1\times ... \times\mathcal{C}_r)\widehat{C}CB)^{\perp}=
((\mathcal{C}_1\times ... \times\mathcal{C}_r)\widehat{C}\widehat{Q}\
\widehat{Q}_t{C}_t^{-1})^{\perp}= $$
$$=((\mathcal{C}_1\times ... \times\mathcal{C}_r)\widehat{Q}\
\widehat{Q}_t\overline{C}{C}_t^{-1})^{\perp}=((\mathcal{C}_1\times ... \times\mathcal{C}_r)CBC_t\overline{C}{C}_t^{-1})^{\perp}=$$
$$=((\mathcal{C}_1\times ... \times\mathcal{C}_r)CB)^{\perp}C^{-1}\overline{C}_t^{-1}C=((\mathcal{C}_1\times ... \times\mathcal{C}_r)C)^{*}C^{-1}\overline{C}_t^{-1}C=$$
$$=(\mathcal{C}_1^*\times ... \times\mathcal{C}_r^*)CC^{-1}\overline{C}_t^{-1}C=(\mathcal{C}_1^*\times ... \times\mathcal{C}_r^*)\overline{C}_t^{-1}C,$$
i.e. $\mathcal{C}^{*}=(\mathcal{C}_1^*\times ... \times\mathcal{C}_r^*)\overline{C}_t^{-1}C$. \hfill $\square$

\medskip

Finally, we have the following

\begin{prop}
Let $\mathcal{C}_T\subseteq\mathbb{F}_q^n$ be a semi-linear
$T$-code invariant under a $\theta$-semi-linear transformation $T=\Theta\circ \overline{M}$. If there exists a matrix $\widehat{M}$ such that
$B_\theta\widehat{M}=\overline{M}B$, then the quasi-Euclidean dual code $\mathcal{C}_T^{*}$ is
a $T'$-code, where $T'=\Theta^{-1}\circ ({{\widehat{M}}_t})_{\theta^{-1}}$.
\end{prop}

\noindent\textit{Proof}. Note that the linear code $\mathcal{C}_TB$ is invariant under the $\theta$-semi-linear transformation $\Theta\circ\widehat{M}$.
Thus we can conclude by Theorem~\ref{prop quasi euclidean}~(i) and Proposition~\ref{euclidean dual proposition}. \hfill $\square$

\subsection{Hermitian duals}

Assume that the order $s$ of $\theta\in Aut(\mathbb{F}_q)$ divides
$m_i$ for every $i=1,...,r$, i.e.
\begin{equation}
m_i= m_i' \cdot s \ ,\ m_i' \in \mathbb{N} \ . \tag{$\diamond\diamond$ }
\end{equation}

\noindent Note that assumption $(\diamond\diamond)$ is always satisfied when $\theta=id$.

Define a ``conjugation'' map $\Phi$ on $R_m:=R/R(X^{m_1}-1)\times
...\times R/R(X^{m_r}-1)$ such that
$$\Phi((a_{i_1}X^{i_1},...,a_{i_r}X^{i_r})):=(\Phi_1(a_{i_1}X^{i_1}),...,\Phi_r(a_{i_r}X^{i_r})),$$
where
$$\Phi_k(a_{i_k}X^{i_k}):=\theta^{-i_k}(a_{i_k})X^{m_k -i_k}\in R/R(X^{m_k}-1)$$ for $k=1,...,r$, which is extended to all elements of
$R_m$ by linearity of addition.

We then define a product of two elements
$\vec{p}(X)=(p_1(X),...,p_r(X))\in R_m$ and
$\vec{t}(X)=(t_1(X),...,t_r(X))\in R_m$ by
$$\vec{p}(X) *_{\widehat{P}} \vec{t}(X):=(p_1(X)\Phi_1 (t_1(X)),...,p_r(X)\Phi_r (t_r(X))).$$

By the above commutative diagram, we can also define a Hermitian
product of two elements $\vec{a}(X):=(a_1(X),...,a_r(X))$ and
$\vec{b}(X):=(b_1(X),...,b_r(X))$ of $R_n:=R/Rf_1\times ...\times
R/Rf_r$ by
$$<\vec{a}(X), \vec{b}(X)>:=j(\vec{a}(X))*_{\widehat{P}} j(\vec{b}(X)).$$

The next two results are now an immediate generalization of
\cite[Proposition 3.2 and Corollary 3.3]{LS}.

\begin{prop}\label{product T-Hermitian dual}
Assume that $(\diamond\diamond)$ holds. Let
$\vec{a}=(\vec{a_1},...,\vec{a_r}),\vec{b}=(\vec{b_1},...,\vec{b_r})\in\mathbb{F}_q^{n_1}\times
... \times \mathbb{F}_q^{n_r}$ and denote by
$$\vec{a}(X):=(\pi_1(\vec{a_1}),...,\pi_r(\vec{a_r})):=(a_1(X),...,a_r(X))$$ and
$$\vec{b}(X):=(\pi_1(\vec{b_1}),...,\pi_r(\vec{b_r})):=(b_1(X),...,b_r(X))$$ their
polynomial representation in $R/Rf_1\times ... \times R/Rf_r$ via
$\pi=(\pi_1,...,\pi_r)$ respectively. If $( \diamond\diamond )$
holds, then
$$\vec{a_i}\cdot_{*_i} \vec{b_i}(\Theta\circ M_i)^{h_i}=0,\ \mathrm{\ for\ all\ } 0\leq h_i\leq m_i-1, \
i=1,...,r\ \iff \ <\vec{a}(X), \vec{b}(X)>=\vec{0}.$$
\end{prop}

\noindent\textit{Proof}. Without loss of generality, we can assume
that $r=1$, since the statement will follow easily by applying the
below argument to each component of $<\vec{a}(X), \vec{b}(X)> \in
R_m$. Moreover, for simplicity we omit the subindexes.

Since $\theta^m=id$, the condition $<a(X), b(X)>=0$ is equivalent
to
\begin{equation*}
\begin{split}
j(a(X))*_{\widehat{P}} j(b(X))=0 & \iff a(X)q_f\Phi (b(X)q_f)=0 \\
 & \iff \left( \sum_{i=0}^{m-1}a'_iX^i\right)\Phi\left(
\sum_{k=0}^{m-1}b'_kX^k\right)=0 \\
& \iff \left( \sum_{i=0}^{m-1}a'_iX^i\right)\left(
\sum_{k=0}^{m-1}\theta^{-k}(b'_k)X^{m-k}\right)=0 \\
 & \iff
\sum_{h=0}^{m-1}\left(
\sum_{i=0}^{m-1}a'_{i+h}\theta^{h}(b'_i)\right)X^h=0,
\end{split}
\end{equation*}
where the subscript $i+h$ is taken modulo $m$. Comparing the
coefficients of $X^h$ on both sides of the last equation, we get
$$\sum_{i=0}^{m-1}a'_{i+h}\theta^{h}(b'_i)=0,\mathrm{\ for\ all\ } 0\leq h\leq m-1.$$
By Proposition \ref{prop} the above equation is equivalent for all
$0\leq h\leq m-1$ to
\begin{equation*}
\begin{split}
\vec{a'}\cdot\vec{b'}(\Theta^h\circ P^h)=0 & \iff \vec{a'}\cdot
\vec{b'}(\Theta\circ P)^h=0
\\
& \iff i(\vec{a})\cdot i(\vec{b})(\Theta\circ P)^h=0 \\
& \iff i(\vec{a})\cdot i(\vec{b}(\Theta\circ M)^h)=0 \\
& \iff \vec{a}Q\cdot (\vec{b}(\Theta\circ M)^h)Q=0,
\end{split}
\end{equation*}
i.e. $\vec{a}\cdot_* \vec{b}(\Theta\circ M)^h=0$ for all $0\leq
h\leq m-1$. \hfill $\square$

\bigskip

Let $I$ be a subset of $R_n$. We define the dual
$I^{<,>}$ of $I$ in $R_n$ taken with respect to the Hermitian
product $<,>$ as
$$I^{<,>}:=\{ \vec{a}(X)\in R_n\ |\ <\vec{a}(X),\vec{t}(X)>=\vec{0}\ ,\  \forall \vec{t}(X)\in I\ \}.$$

\begin{defn}\label{definition bis}
Let $T$ be a semi-linear transformation of $\mathbb{F}_q^n$ as in $(*)$.
We define the \textit{Hermitian dual code} $\mathcal{C}^{\nu}$ of a linear code $\mathcal{C}\subseteq\mathbb{F}_q^n$ with respect to
$<,>$ as the linear code
$$\mathcal{C}^{\nu}:=\left\{\vec{x}\in\mathbb{F}_q^n\ |\ <\vec{x}(X),\vec{c}(X)>=0\ \mathrm{for\ every}\ \vec{c}\in \mathcal{C}\right\}.$$
\end{defn}

\begin{rem}
If $I\subseteq R_n$ is a left $R$-submodule, then $I^{<,>}$ is again a left $R$-submodule of $R_n$. Consequently,
from Theorem \ref{f-module inv} we can deduce that if $\mathcal{C}'$ is a code invariant under $D$ as in $(*)$,
then ${\mathcal{C}'}^{\nu}=\pi^{-1}(\pi(\mathcal{C}')^{<,>})$ is again a code invariant by $D$.
\end{rem}

\medskip

From Proposition \ref{product T-Hermitian dual} we can
deduce the following results which relate the quasi-Euclidean duals with the Hermitian dual codes of product $T$-codes.

\begin{thm}\label{T-cor duals}
Let $\mathcal{C}_T=(\mathcal{C}_1\times ...
\times\mathcal{C}_r)C$ be a product $T$-code and define the isomorphism
$\overline{\pi}=\pi\circ\varphi$. If $( \diamond\diamond )$ holds,
then
$$\overline{\pi} (\mathcal{C}_T^{*})=\overline{\pi} (\mathcal{C}_T)^{<,>}\ .$$
\end{thm}

\noindent\textit{Proof}. Since $\mathcal{C}_T=(\mathcal{C}_1\times
... \times\mathcal{C}_r)C$, from Corollary \ref{T-cor} we deduce that
$\mathcal{C}_T^{*}=(\mathcal{C}_1^*\times ...
\times\mathcal{C}_r^*)C$. Thus it is sufficient to prove that
$$\pi (\mathcal{C}_1^*\times ... \times\mathcal{C}_r^*)=\pi (\mathcal{C}_1\times ... \times\mathcal{C}_r)^{<,>}\ .$$
Moreover, without loss of generality, we can assume that $r=1$.
Therefore, let $\pi (\vec{b})=\pi_1 (\vec{b})\in\pi_1 (\mathcal{C}_1^*)$ for some $\vec{b}\in
\mathcal{C}_1^*$. Then for every $\vec{a}\in \mathcal{C}_1$ and
$h\in\mathbb{Z}_{\geq 0}$ we have
$\vec{b}\cdot_* \vec{a}(\Theta\circ M_1)^h=0.$ Thus by Proposition \ref{product T-Hermitian dual} we get
$< \pi_1 (\vec{b}),\pi_1 (\vec{a})>=0\ \mathrm{\ for\ all\ } \vec{a}\in \mathcal{C}_1,$ i.e.
$\pi_1 (\vec{b})\in\pi_1 (\mathcal{C}_1)^{<,>}$. Hence $\pi_1 (\mathcal{C}_1^*)\subseteq \pi_1 (\mathcal{C}_1)^{<,>}$.
Finally, let $b(X)\in\pi_1(\mathcal{C}_1)^{<,>}$. Then we get
$<b(X),\pi_1 (\vec{a})>=0,\ \forall \vec{a}\in \mathcal{C}_1$. By Proposition \ref{product T-Hermitian dual} with $h=0$, this implies that
$\pi_1^{-1}(b(X))\cdot_*\vec{a}=0,\ \forall \vec{a}\in \mathcal{C}_1$, i.e. $\pi_1^{-1}(b(X))\in \mathcal{C}_1^*$. This shows that
$b(X)=\pi_1(\pi_1^{-1}(b(X)))\in \pi_1(\mathcal{C}_1^*),$ that is, $\pi_1 (\mathcal{C}_1)^{<,>}\subseteq\pi_1 (\mathcal{C}_1^*)$. \hfill $\square$

\begin{rem}
Let $\mathcal{C}_T=(\mathcal{C}_1\times ...
\times\mathcal{C}_r)C$ be a product $T$-code. Then from Theorem \ref{T-cor duals} we deduce that $\varphi(\mathcal{C}^*)=(\varphi (\mathcal{C}))^\nu$.
In particular, if $\mathcal{C}=\mathcal{C}_1\times ...
\times\mathcal{C}_r$ is a product of module $\theta$-codes $\mathcal{C}_i$ for $i=1,...,r$, then we have
$\mathcal{C}^*=\mathcal{C}^\nu$.
\end{rem}

\begin{cor}\label{cor quasi-euclidean and hermitian}
Let $\mathcal{C}_T=(\mathcal{C}_1\times ...
\times\mathcal{C}_r)C$ be a product $T$-code, where
$C$ is as in $(*)$. If $( \diamond\diamond )$ holds, then
$$\mathcal{C}_T=\mathcal{C}_T^{*} \iff \overline{\pi} (\mathcal{C}_T)=\overline{\pi} (\mathcal{C}_T)^{<,>}\ ,$$ i.e., \ $\mathcal{C}_T$ is self-dual with respect to
$\cdot_{*} \iff \overline{\pi} (\mathcal{C}_T)$ is self-dual with
respect to $<,>$.
\end{cor}

\noindent\textit{Proof}. Since $\overline{\pi}$ is an isomorphism,
this follows immediately from Theorem \ref{T-cor duals}. \hfill
$\square$

\begin{thm}\label{T-theorem quasi-euclidean}
Let $\mathcal{C}_T=(\mathcal{C}_1\times ...
\times\mathcal{C}_r)C$ be a product $T$-code, where $\mathcal{C}_i=(g_i)_{n_i,q}^{k_i,\theta}$ is an $f_i$-module $\theta$-codes for every $i=1,...,r$.
If $( \diamond\diamond )$ holds, then
$\mathcal{C}_T^{*}=(\mathcal{C}_1^*\times ... \times\mathcal{C}_r^*)C$
is a product $T$-code, where
$\mathcal{C}_i^*=(\mathcal{C}_iB_i)^{\perp}$ with
$B_i:=Q_i(Q_i)_t$ for $i=1,...,r$. Furthermore, a generator matrix
for $\mathcal{C}_T^{*}$ is given by
$$G^{*}:=\left(\begin{array}{cccc}
G_1^* &  &  &  \\
 & G_2^* &  &  \\
 &  & \ddots &  \\
 &  &  & G_r^*
\end{array}\right)\cdot C\ ,$$
where
$$G_i^*:=\left(\begin{array}{c}
\pi_i^{-1}(g_i^*)  \\ [2pt]
\pi_i^{-1}(g_i^*)(\Theta\circ M_i)  \\  [2pt]
 \vdots  \\  [2pt]
\pi_i^{-1}(g_i^*)(\Theta\circ M_i)^{s_i-1}  \\  [2pt]
\end{array}\right)\ , $$
with $s_i:=\dim\mathcal{C}_i^*$, $g_i^*q_{f_i}=l.l.c.m(h_i^{\perp},q_i)\mod (X^m_i-1)$, $h_i^{\perp}=\sum_{j=0}^{k_i}\theta^i(h_{k_i-j})X^j$ and
$X^{m_i}-1=g_iq_{f_i}(\sum_{j=0}^{k_i}h_jX^j)$, is the generator matrix of the quasi-Euclidean code $\mathcal{C}_i^*$ for
every $i=1,...,r$.
\end{thm}

\noindent\textit{Proof}. Since $\mathcal{C}_T=(\mathcal{C}_1\times ...
\times\mathcal{C}_r)C$ is a product $T$-code, then $\overline{\pi} (\mathcal{C}_T)$ is
a left $R$-submodule of $R_n$. Hence $\overline{\pi} (\mathcal{C}_T)^{<,>}$ is a left $R$-submodule.
By Proposition \ref{T-cor} and Theorems \ref{T-cor duals} and \ref{f-module inv}, we conclude that
$\mathcal{C}_T^{*}=(\mathcal{C}_1^*\times ... \times\mathcal{C}_r^*)C$
is a product $T$-code.

Consider the following commutative diagrams
\begin{displaymath}
    \xymatrix{
        \mathcal{C}_k \ar[r]^{i_k} \ar[d]_{\pi_k} &  i_k(\mathcal{C}_k) \ar[d]^{\pi_k'} \\
            (g_k)  \ar[r]_{j_k}   & (g_kq_{f_k}) } \qquad
            \xymatrix{
        \mathcal{C}_k^* \ar[r]^{i_k} \ar[d]_{\pi_k} &  i_k(\mathcal{C}_k^*) \ar[d]^{\pi_k'} \\
            (g_k^*)  \ar[r]_{j_k}   & (G_k) }
    \end{displaymath}
for every $k=1,...,r$. By Proposition \ref{prop} we see that $i_k(\mathcal{C}_k)$ is a $\theta$-cyclic code.
So from \cite[Theorem 8]{BU3} we know that $i_k(\mathcal{C}_k)^{\perp}$ is again a $\theta$-cyclic code
generated by the skew polynomial $h_k^{\perp}:=h_k^*\in R$ such that $X^{m_k}-1=g_kq_{f_k}{h_k}$, where
$h^*$ is as in \cite[Definition 3]{BU1}. Since $\mathcal{I}_k:=\mathrm{Im\ }i_k$ is generated by $q_{f_k}\in R$,
from Theorem \ref{prop quasi euclidean} (v) it follows that $\pi_k'(i_k(\mathcal{C}_k^*))=(h_k^{\perp})\cap (q_{f_k})$,
i.e. $\pi_k'(i_k(\mathcal{C}_k^*))=(G_k)$ with $G_k=l.l.c.m.(h_k^{\perp},q_{f_k})$.
From Proposition \ref{lemma} we deduce that $\pi_k(\mathcal{C}_k^*)=(g_k^*)$
with $g_k^*$ such that $G_k=g_k^* q_{f_k}$.
\hfill $\square$

\begin{cor}\label{C+Ker B}
Let $\mathcal{C}=(\mathcal{C}_1\times ...
\times\mathcal{C}_r)C$ be a product $T$-code, where $C$ is as in $(*)$.
If $( \diamond\diamond )$ holds, then

\smallskip

\qquad $\mathcal{C}^*$ is a product $T$-code $\iff$ $\mathcal{C}+\mathrm{Ker}\ B$ is a product $T$-code.

\end{cor}

\noindent\textit{Proof}. Suppose that $\mathcal{C}^*=(\mathcal{C}_1^*\times ...
\times\mathcal{C}_r^*)C$ is a product $T$-code. Then by Proposition \ref{prop quasi euclidean} (iv)
and Corollary \ref{T-theorem quasi-euclidean} we see that $\mathcal{C}+\mathrm{Ker}\ B=(\mathcal{C}^*)^*$ is a
product $T$-code. Finally, assume that $\mathcal{C}+\mathrm{Ker}\ B$ is a product $T$-code. Then by Theorem
\ref{prop quasi euclidean} (vi) and Corollary \ref{T-theorem quasi-euclidean}, we deduce that
$\mathcal{C}^*=\mathcal{C}^*\cap (\mathrm{Ker}\ B)^*=(\mathcal{C}+\mathrm{Ker}\ B)^*$ is a product $T$-code.
\hfill $\square$

\medskip

Let us note here that the converse of Corollary \ref{T-theorem quasi-euclidean} is not true in general, as the
following example shows.

\begin{ex}
In $\mathbb{F}_{4}^3$, where $\mathbb{F}_{4}=\mathbb{F}_{2}[\alpha]$ with $\alpha^2+\alpha+1=0$, consider the polynomial
$f_2=X^3+X^2+\alpha X+\alpha^2$. Then from Example \ref{example3} we know that $m=6$ and
$$B_2=\left(
\begin{array}{ccc}
0 & \alpha^2 & 0 \\
\alpha^2 & 0 & \alpha \\
0 & \alpha & 0
\end{array}\right),$$
with $\mathrm{rk\ }B_2=2$. Consider the linear code $\mathcal{C}\subset\mathbb{F}_{4}^3$ generated by the vectors $\vec{e}_2=(0,1,0)$ and $\vec{e}_3=(0,0,1)$.
Since
$$(\vec{e}_3)\ \Theta\circ\left(
\begin{array}{ccc}
0 & 1 & 0 \\
0 & 0 & 1 \\
\alpha^2 & \alpha & 1
\end{array}\right)=\vec{e}_3\left(
\begin{array}{ccc}
0 & 1 & 0 \\
0 & 0 & 1 \\
\alpha^2 & \alpha & 1
\end{array}\right)=(\alpha^2,\alpha,1)\notin \mathcal{C},$$ we see that $\mathcal{C}$ is not an $f_2$-module $\theta$-code.
On the other hand, since $\mathrm{Ker}\ B_2$ is generated by the vector $(\alpha^2,0,1)$ and $\mathcal{C}\cap\mathrm{Ker}\ B_2=\{\vec{0}\}$,
we obtain that $$\mathcal{C}+\mathrm{Ker}\ B_2=\mathcal{C}\oplus\mathrm{Ker}\ B_2=\mathbb{F}_4^3$$ is an $f_2$-module $\theta$-code.
By Corollary \ref{C+Ker B} we get that $\mathcal{C}^*$ is an $f_2$-module $\theta$-code.
\end{ex}

\begin{rem}
If $( \diamond\diamond )$ holds, then $\mathrm{Ker\ }B\subseteq\mathbb{F}_q^n$ is a $T$-code such that $\mathrm{Ker\ }B=(\mathrm{Ker\ }B_1\times ...\times\mathrm{Ker\ }B_r)$,
\ $\mathrm{Ker\ }B^{\perp}=\mathrm{Im\ }B,\ (\mathrm{Ker\ }B^{\perp})^{\perp}=\mathrm{Ker\ }B$ and
$\mathrm{Ker\ }B^{*}=\mathbb{F}_q^n,\ (\mathrm{Ker\ }B^{*})^{*}=\mathrm{Ker\ }B$.
In particular, $\mathrm{Ker\ }B\subseteq\mathbb{F}_q^n$ does not contain any
$T$-code $\mathcal{C}\subseteq\mathbb{F}_q^n$ with $\mathcal{C}^*\neq\mathbb{F}_q^n$.
\end{rem}

\section{An encoding and decoding algorithm}

Given a $\theta$-semi-linear transformation $T=\Theta\circ M$ and
a product $T$-code $\mathcal{C}_T\subset\mathbb{F}_q^n$ of dimension $k<n$, a
classical codification of a message $\vec{M}\in\mathbb{F}_q^k$ is given by $\vec{M}G_T{C}^{-1}$, where
$G_T$ is a generator matrix of $\mathcal{C}_T$ and $C$ is the invertible matrix such that
$\mathcal{C}_T=(\mathcal{C}_1\times ... \times \mathcal{C}_r)C$.
Note that $\vec{M}G_T\in \mathcal{C}_T$ and
$$\vec{m}:=\vec{M}G_T{C}^{-1}\in \mathcal{C}_T{C}^{-1}=\mathcal{C}_1\times ... \times \mathcal{C}_r$$
for some $f_i$-module $\theta$-codes $\mathcal{C}_i=(g_i)$, where the $g_i$'s are right divisors of the $f_i$'s respectively
(see Remark \ref{semilinear-rem} and assumption (*)). However, this encoding method is not systematic, i.e. it is not
strictly related with an easy decoding algorithm.

\medskip

So, let us give here a non-trivial and systematic encoding method for product $T$-codes.
Let $\vec{M}\in\mathbb{F}_q^k=\mathbb{F}_q^{k_1}\times ...\times\mathbb{F}_q^{k_r}$ be the original message such that
$\vec{M}=(\vec{M}_1,...,\vec{M}_r)$, where $\vec{M}_i\in\mathbb{F}_q^{k_i}$ for every $i=1,...,r$.
Let $\mathcal{C}_T=(\mathcal{C}_1\times ...\times\mathcal{C}_r)C$ be a product $T$-code such that $\dim_{\mathbb{F}_q}\mathcal{C}_i=k_i$
for any $i=1,...,r$. Note that $\mathcal{C}_i\subseteq\mathbb{F}_q^{n_i}$ with $n_i\geq k_i$ for every $i=1,...,r$.
Therefore, consider the natural injective
map $i_j:\mathbb{F}_q^{k_j}\to\mathbb{F}_q^{n_j}$ such that $i_j(a_1,...,a_{k_j}):=(a_1,...,a_{k_j},0,...,0)$ for any $j=1,...,r$, and define
the injective map
$$i:=(i_1,...,i_r): \mathbb{F}_q^{k_1}\times ...\times\mathbb{F}_q^{k_r}\to\mathbb{F}_q^{n_1}\times ...\times\mathbb{F}_q^{n_r}.$$
Define $\vec{m}:=i(\vec{M})=((\vec{M}_{1},\vec{0}),\ldots ,(\vec{M}_{r},\vec{0}))\in\mathbb{F}_q^{n_1}\times ...\times\mathbb{F}_q^{n_r}$ and
denote by $m=(m_1,\ldots ,m_r) \in R_n$ the representation of the message $\vec{m}=i(\vec{M})\in\mathbb{F}_q^n=
\mathbb{F}_q^{n_1}\times ...\times\mathbb{F}_q^{n_r}$, via the vector isomorphism
$$\pi:=(\pi_1,...,\pi_r):\mathbb{F}_q^{n_1}\times ...\times\mathbb{F}_q^{n_r}\to R_n:=R/Rf_1\times ... \times R/Rf_r\ .$$

\noindent At this point, we can encode the original message $\vec{m}:=i(\vec{M})$ by working
equivalently on either $(i)$ $R_n$, or $(ii)$
$\mathbb{F}_q^{n}:=\mathbb{F}_q^{n_1}\times ...\times\mathbb{F}_q^{n_r}$.

\medskip

\noindent $(i)$ Multiply the original messages
$m_i$ by $X^{n_i-k_i}$, where $m_i=m_{i,0}+m_{i,1}X+...+m_{i,k_i-1}X^{k_i-1}$ and $k_i=\dim_{\mathbb{F}_q}\mathcal{C}_i$.
The result is $X^{n_i-k_i}\cdot m_i=\theta^{n_i-k_i}(m_{i,0})X^{n_i-k_i}+
\theta^{n_i-k_i}(m_{i,1})X^{n_i-k_i+1}+...+\theta^{n_i-k_i}(m_{i,k_i-1})X^{n_i-1}$ for $i=1,\ldots ,r$.
Write $X^{n_i-k_i}\cdot m_i=q_ig_i+r_i$ for every $i=1,...,r$,
where $\deg r_i<n_i-k_i$. Since $q_ig_i\in \mathcal{C}_i$, we can encode the original message $\vec{m}\in\mathbb{F}_q^n$
by $$\vec{m}':=(\pi_1^{-1}(X^{n_1-k_1}\cdot m_1-r_1),...,\pi_r^{-1}(X^{n_r-k_r}\cdot m_r-r_r))\in\mathcal{C}_1\times ... \times \mathcal{C}_r.$$
Since $\deg r_i<n_i-k_i$ for every $i=1,...,r$, observe that all
the information about the original messages $m_i$ is contained in the last powers
$X^{n_i-k_i}, ... , X^{n_i-1}$ of $X^{n_i-k_i}\cdot m_i-r_i\in \pi_i(\mathcal{C}_i)$.

\smallskip

\noindent $(ii)$ Define the map
$$\overline{\Theta }:\ \begin{array}{ccc}
\mathbb{F} _q ^{n_1}\times  \cdots \times \mathbb{F} _q ^{n_r}&\longrightarrow & \mathbb{F} _q ^{n_1}\times  \cdots \times \mathbb{F} _q ^{n_r}\\
(\vec{x}_1,\ldots ,\vec{x}_r )&\longmapsto &(\vec{x}_1(\Theta \circ M_1)^{n_1-k_1},\ldots ,\vec{x}_r(\Theta \circ M_r)^{n_r-k_r} )
\end{array},$$ where the $M_i$'s are matrices as in Theorem \ref{semilinear}.
By applying $\overline{\Theta }$ to $\vec{m}$ we have
\begin{equation*}
\begin{split}
\vec{m}\overline{\Theta }&= ((\vec{M}_{1},\vec{0})(\Theta \circ M_1)^{n_1-k_1},\ldots ,(\vec{M}_{r},\vec{0})(\Theta \circ M_r)^{n_r-k_r} )\\
&=((\vec{0},(\vec{M}_{1})\Theta ^{n_1-k_1}),\ldots ,(\vec{0},(\vec{M}_{r})\Theta ^{n_r-k_r}) )
\end{split}
\end{equation*}
If $\vec{m}':=((\vec{c}_1,(\vec{M}_{1})\Theta ^{n_1-k_1}),\ldots ,(\vec{c}_r,(\vec{M}_{r})\Theta ^{n_r-k_r}))$ is such that $\vec{m}'H_t=\vec{0}$,
where
$$H=\left(
\begin{array}{ccc}
H_1&&\\
&\ddots &\\
&&H_r
\end{array} \right)$$ is the parity check matrix of $\mathcal{C}_1\times ... \times \mathcal{C}_r$
and the matrices $H_i=(I_{n_i-k_i}\ |\ (T_i)_t)$ are given by Proposition \ref{pc matrix} for every $i=1,...,r$. Then
$\vec{m}' \in \mathcal{C}_1\times \cdots \times \mathcal{C}_r$ is the encoded message of $\vec{m}\in\mathbb{F}_q^n$.

\medskip

Now, let $\vec{m}''$ be the received message. If during the transmission of the encoded message $\vec{m}'$ there were not errors,
i.e. $\vec{m}'' \in \mathcal{C}_1\times \cdots \times \mathcal{C}_r$, then in both cases $(i)$ and $(ii)$ we can decode $\vec{m}''=(\vec{m}_1'',...,\vec{m}_r'')$
by applying $\Theta^{-n_i+k_i}$ to each component $\vec{m}_i''$ of $\vec{m}''$.
The original components $\vec{m}_i$ of $\vec{m}=(\vec{m}_1,...,\vec{m}_r)$ will be given by the last $k_i$ coordinates of
$(\vec{m}_i'')\Theta^{-n_i+k_i}$ for every $i=1,...,r$.

\medskip

Finally, if there were errors during the transmission of the message $\vec{m}'$, i.e. $\vec{m}'' \notin \mathcal{C}_1\times \cdots \times \mathcal{C}_r$, then by assuming that the
error $\vec{e}$, defined as
$$\vec{e}:=\vec{m}''-\vec{m}'\in \vec{m}''+(\mathcal{C}_1\times ... \times \mathcal{C}_r),$$
where $\vec{m}''$ and $\vec{m}'$ are the received and the encoded messages
respectively, has small weight $wt(\vec{e})$, we can use the below
error detecting and correcting algorithm inspired by \cite{HP} and then the above decoding procedure.

\newpage

\noindent\textbf{A Meggitt type error correcting algorithm.}

\smallskip

Put $d_{min}:=\min_{i=1,...,r}\{d(\mathcal{C}_i)\}$, where $d(\mathcal{C}_i):=d_i$ is the minimum Hamming distance
of the code $\mathcal{C}_i$ for $i=1,...,r$, and assume that
$$wt(\vec{e})\leq\frac{d_{min}-1}{2}.$$
Let $\pi_j:\mathbb{F}_q^{n_j}\to R/Rf_j$ be the usual isomorphism for every $j=1,...,r$.

For any vector
$\vec{v}=(\vec{v}_1,...,\vec{v}_r)\in \mathbb{F}_q^{n}=\mathbb{F}_q^{n_1}\times ...\times\mathbb{F}_q^{n_r}$ put
$\pi (\vec{v}):=(\pi_1(\vec{v}_1),...,\pi_r(\vec{v}_r))$ and define the syndrome of $\pi (\vec{v})$ as follows:
$$S(\pi (\vec{v})):=(R_{g_1}(\pi_1(\vec{v}_1)),...,R_{g_r}(\pi_r(\vec{v}_r))),$$
where $R_{g_i}(\pi_i(\vec{v}_i))$ is the rest of the division of $\pi_i(\vec{v}_i)$ by $g_i$ for every $i=1,...,r$.

Observe that $S(\pi (\vec{m}'))=(0,...,0)$. Hence $S(\pi (\vec{e}))=S(\pi (\vec{m}''))$.
Denote by $t_i$ the polynomials in $R$ such that $t_i\cdot X=1$ in $R/Rf_i$ for every $i=1,...,r$.

\bigskip

\noindent\textbf{Algorithm 2:}

\smallskip

\textbf{Input}: $\vec{m}''=(\vec{m}_1'',...,\vec{m}_r'')$
\begin{itemize}
\item {\bf Step 1}: Compute all the syndromes
$$S(\pi(\vec{e'}))=S((\pi_1(\vec{e'}_1),...,\pi_r(\vec{e'}_r)),$$ where $\pi_i(\vec{e'}_i)=\sum_{j=0}^{n_i-1} e_{j,i}'X^j$ is such that
$wt(\vec{e'}_i)=wt(\pi_i(\vec{e'}_i))\leq\frac{d_i-1}{2}$ with $d_i$ the minimum Hamming distance of $\mathcal{C}_i$ and $e_{i,n_i-1}'\neq 0$;
\item {\bf Step 2}: Compute $S(\pi(\vec{m}''))$ and define $\vec{s}:=S(\pi(\vec{m}''))$;
\item {\bf Step 3}: If $\vec{s}=\vec{0}\in R/Rf_1\times ... \times R/Rf_r$ then write $\vec{e}=\vec{0}$;
\item {\bf Step 4}: If $\vec{s}$ is equal to some of the syndromes $S(\pi(\vec{e'}))$ of Step 1, then write $\vec{e}=\vec{e'}$;
\item {\bf Step 5}: If $\vec{s}$ is not in the list of Step 1, then
$$\vec{m}''=\vec{m}'+\vec{e''}$$ for some error $\vec{e''}=(\vec{e''}_1,...,\vec{e''}_r)\in \mathbb{F}_q^{n_1}\times ...\times\mathbb{F}_q^{n_r}$
such that $wt(\vec{e''})\leq\frac{d_{min}-1}{2}$ and
$\pi(\vec{e''})=(\sum_{j=0}^{h_1} e^{''}_{j,1}X^j,...,\sum_{j=0}^{h_r} e^{''}_{j,r}X^j)$ with $e^{''}_{h_i,i}\neq 0$, $h_i\leq n_i-1$ and $h_k< n_k-1$
for some $k=1,...,r$. Since $\theta$ is an automorphism of $\mathbb{F}_q$, there exists an integer
$\delta_k:=n_k-h_k-1$ such that
$$\overline{e}_k:=X^{\delta_k}\cdot \left(\sum_{j=0}^{h_k} e^{''}_{j,k}X^j\right),$$
i.e. $\pi_k(\vec{E}_k):=\overline{e}_k=\sum_{j=0}^{n_k-1} \overline{e}_{j,k}X^j$ is such that
$wt(\vec{E}_k)\leq\frac{d_k-1}{2}$
and $\overline{e}_{n_k-1,k}\neq 0$.
Thus the syndrome
$$S((\pi_1(\vec{e'}_1),..., \overline{e}_k,...,\pi_r(\vec{e'}_r))$$
is as in Step 1, where $\pi_i(\vec{e'}_i)=\sum_{j=0}^{n_i-1} {e'}_{j,i}X^j$ is such that
$wt(\pi_i(\vec{{e'}}_i))\leq\frac{d_i-1}{2}$ and ${e}_{i,n_i-1}'\neq 0$ for $i\neq k$.
Then write
$$\vec{e}=(\vec{e'}_1,..., \pi_k^{-1}(t_k^{\delta_k}\cdot\overline{e}_k),...,\vec{e'}_r);$$
\end{itemize}

\textbf{Output}: $\vec{m}'=\vec{m}''-\vec{e}$.

\section{A method to construct $T$-codes}

Observe that to construct a product $T$-code (see Definition \ref{product semilinear codes}) it is sufficient
to construct module $\theta$-codes (see Definition \ref{module skew codes}).

Note that in $R$ there are exactly $q^{r-1}(q-1)$ different
polynomials of the form $g=g_0+g_1X+...+g_{r-1}X^{r-1}+X^r$ with $g_0\neq 0$. Thus if $h$ is another
monic polynomial of degree $r$, then it follows that $(g)\neq (h)$ whenever $g\neq h$. Furthermore, for any given monic
polynomial $g\in R$ of degree $r<n$ as above there exists a polynomial $f\in R$ of degree $n$ such that $g$ is a (right)
divisor of $f$. This shows that there exist $q^{r-1}(q-1)$ module $\theta$-codes with parameters of $[n,n-r]_q$.

From now on, a linear code $\mathcal{C}$ of type $[n,k]_q$ with Hamming distance equal to $d$ will be called simply a code of type $[n,k,d]_q$.

So, let us give here the following

\begin{defn}
$$D_q^{\theta}(n,k):=\max\left\{d\ |\ \exists\mathrm{\ a\ module\ } \theta\mathrm{-code\ of\ type\ } [n,k,d]_q \right\}$$
\end{defn}

Similarly to \cite[Proposition 3.1]{LL}, we can obtain the following

\begin{prop}\label{construction distance}
$$D_q^{\theta}(n,k)\geq D_q^{\theta}(n+1,k+1) .$$
\end{prop}

\noindent\textit{Proof}.
Let $g=g_0+g_1X+...+g_{n-k}X^{n-k}$ be the generator polynomial of a module $\theta$-code $\mathcal{C}_{n+1,k+1}$
with parameters $[n+1,k+1,D_q^{\theta}(n+1,k+1)]$. Observe that $g_0$ and $g_{n-k}$ are distinct to zero and that the
generator matrix $G_{n+1,k+1}$ of $\mathcal{C}_{n+1,k+1}$ has the form
$$\left(
\begin{array}{c|cccccc}
g_0 & g_1 & ... & g_{n-k} & 0 & ... & 0 \\
\hline
0 &  &  &  &   &  &  \\
 \vdots &  &  & G_{n,k} &  &  &  \\
0 &  & &  &  & &
\end{array}\right) ,$$
where $G_{n,k}$ is the following matrix
$$\left(
\begin{array}{ccccccc}
\theta(g_{0}) & ... & \theta(g_{n-k}) & 0 & ... & 0 \\
0 & \theta^2(g_{0}) & ... & \theta^2(g_{n-k}) & ... & 0 \\
\vdots &   & \ddots & \ddots &  & \vdots \\
0 & ... & 0 & \theta^k(g_{0}) & ... & \theta^k(g_{n-k})
\end{array}\right) .$$
Note that the minimum (Hamming) distance decided by $G_{n,k}$ is at least
$D_q^{\theta}(n+1,k+1)$. Define $G:=\theta(g_{0})+\theta(g_{1})X ... + \theta(g_{n-k})X^{n-k}$.
Then $G$ is the generator polynomial of a module $\theta$-code $\mathcal{C}_{n,k}$ of type $[n,k,d]_q$ with
$d\geq D_q^{\theta}(n+1,k+1)$. Hence we get $D_q^{\theta}(n,k)\geq d\geq D_q^{\theta}(n+1,k+1)$.
\hfill $\square$

\begin{rem}
If $\mathcal{C}$ is a module $\theta$-code of type $[n,k,\Delta]_q$ with distance $\Delta\geq 1$, then we have $D_q^{\theta}(n,k)\geq\Delta$.
Therefore by Proposition \ref{construction distance} we see that for any integer $\delta$ such that $0\leq\delta<k$ there exists
at least a module $\theta$-code $\mathcal{C}'$ of type $[n-\delta,k-\delta,d]_q$ with $d\geq\Delta$. Thus the above result can be useful to
ensure the existence and the construction of module $\theta$-codes of type $[n,k,d]_q$ with distance $d$ greater than or equal to some fixed value $\Delta$ and small values for $n$ and $k$.
\end{rem}

Denote by $\mathbb{F}_q^{\theta}\subseteq\mathbb{F}_q$ the field fixed by $\theta$. In what follows we try to construct
vectors $\vec{v}\in \mathbb{F}_q^n$ such that $1\leq\dim [\vec{v}]\leq k$ for some integer $k<n$, where $ [\vec{v}]\subset \mathbb{F}_q^n$
is the vector subspace generated by $\left\{ \vec{v}, (\vec{v})(\Theta\circ A_c),(\vec{v})(\Theta\circ A_c)^2,...\right\}$ and $A_c$ is
the companion matrix of $f\in R$ as in Remark \ref{m identity case}.

For simplicity, put $A:=A_c$ and note that
$$(\vec{v})(\Theta\circ A_{\theta})=(\vec{v})(A\circ \Theta)$$
for any $\vec{v}\in (\mathbb{F}_q)^n$, where $A_{\theta}:=[\theta (a_{ij})]$ if $A=[a_{ij}]$. This gives the following

\begin{lem}\label{powers}
For every integer $k\geq 1$, we have
$$(\Theta \circ A)^k=\Theta^k\circ A_k,$$ where $A_k:= A_{\theta^{k-1}}\cdot ...\cdot A_{\theta^2}\cdot A_{\theta}\cdot A$
for $k\geq 2$ and $A_1:=A$.
\end{lem}

Let $h$ be an integer such that $1\leq h\leq n-1$ and consider the equation:
$$(\#)\qquad (\vec{v})(\Theta\circ A)^hx_h+...+(\vec{v})(\Theta\circ A)^1x_1+(\vec{v})x_0=\vec{0}.$$
If there exists a non-trivial vector $\vec{v}$ and a non-zero $x_h\in\mathbb{F}_q$ which satisfies the above equation $(\#)$,
we can deduce that $(\vec{v})(\Theta\circ A)^h$ can be written as a linear combination of vectors in
$\left\{ \vec{v}, (\vec{v})(\Theta\circ A),...,(\vec{v})(\Theta\circ A)^{h-1} \right\}$, i.e. $1\leq\dim [\vec{v}]\leq h$.

In order to simplify equation $(\#)$, we will consider only vectors $\vec{v}\in (\mathbb{F}_q^{\theta})^n$.
In this case, by Lemma \ref{powers}
$(\#)$ becomes
$$(\#')\qquad \vec{v}\cdot (A_hx_h+...+A_1x_1+Ix_0)=\vec{0},$$
where $\vec{v}\in (\mathbb{F}_q^{\theta})^n$. Thus the existence of a non-trivial vectors $\vec{v}\in (\mathbb{F}_q^{\theta})^n$
which satisfy equation $(\#')$ implies the existence of non-trivial solutions $x_h,...,x_1,x_0$ of the equation
$$(\#'')\qquad\det (A_hx_h+...+A_1x_1+Ix_0)=0.$$ So we can translate the problem of finding a vector $\vec{v}\neq\vec{0}$ in $(\mathbb{F}_q^{\theta})^n$
which is a solution of $(\#)$ to the problem of finding non-trivial solutions $x_h,...,x_1,x_0$ in $\mathbb{F}_q$ of $(\#'')$.
Define
$$F_h(x_0,x_1,...,x_h):=\det (A_hx_h+...+A_1x_1+Ix_0).$$ We have the following

\begin{lem}\label{homogeneous}
The polynomial $F_h(x_0,x_1,...,x_h)$ is an homogeneous polynomial of degree $n$ in
the variables $x_0,x_1,...,x_h$.
\end{lem}

\noindent\textit{Proof}.
For any $\lambda\in\mathbb{F}_q$, we get
\begin{equation*}
\begin{split}
F_h(\lambda x_0,\lambda x_1,...,\lambda x_h) & = \det (A_h(\lambda x_h)+...+A_1(\lambda x_1)+I(\lambda x_0)) \\
& =\det (\lambda I)\cdot \det(A_hx_h+...+A_1x_1+Ix_0) \\
& =\lambda^n\cdot F_h(x_0,x_1,...,x_h),
\end{split}
\end{equation*}
and this gives the statement.
\hfill $\square$

\bigskip

\noindent From Lemma \ref{homogeneous} it follows that the zero locus $Z(F_h(x_0,x_1,...,x_h))$ of $F_h(x_0,x_1,...,x_h)$ on the projective space
$\mathbb{P}^{h}({\mathbb{F}_q})$ is well defined. Put
$$Z_{h,n}:=Z(F_h(x_0,x_1,...,x_h))\subset \mathbb{P}^{h}({\mathbb{F}_q}).$$
Then $Z_{h,n}$ is a hypersurface of $\mathbb{P}^{h}({\mathbb{F}_q})$, i.e. $\dim Z_h=h-1$,
of degree $n\geq h+1$. Moreover, all the points of $Z_{h,n}$ represent no trivial solutions of
$(\#'')$. This gives a
relation between the construction of a module $\theta$-code $\mathcal{C}=[\vec{v}]$ of dimension less or equal to $h$, where
$\vec{v}\in (\mathbb{F}_q^{\theta})^n\cap\mathrm{Ker}\ (A_hx_h+...+A_1x_1+Ix_0)$, with the existence of (rational) points
on the hypersurface $Z_{h,n}$ of $\mathbb{P}^{h}({\mathbb{F}_q})$.

\begin{rem}
The above procedure gives a method to construct an $f$-module $\theta$-code. Moreover, when $\theta=id$
we know from \cite{HK} that the number $N_q$ of $\mathbb{F}_q$-points of the hypersurface $Z_{h,n}$ is bounded for the following inequalities:

(i) $N_q\leq (n-1)q+1$ if $h=2$, except for a curve $Z_{2,4}$ over $\mathbb{F}_4$;

(ii) $N_q\leq (n-1)q^{h-1}+nq^{h-2}+\frac{q^{h-2}-1}{q-1}$ if $h\geq 3$.
\end{rem}

\noindent For the general case of $T$-codes, an argument similar to the above can be directly applied to a semi-linear transformation
$D:=\Theta\circ\mathrm{diag}(M_1,...,M_r)$ instead of
$\Theta\circ A_c$. Recall that any $T$-code $\mathcal{C}_T$ can be obtained from a code $\mathcal{C}_D$ invariant under
$D$ by the relation $\mathcal{C}_T=\mathcal{C}_DC$, where $C$ is an invertible matrix such that
$CTC^{-1}=D$. Therefore, to obtain a $T$-code it is sufficient to construct a code $\mathcal{C}_D$ invariant under
$D$. As above, this allows us to find (rational) solutions of the following equation
$$(\#\#)\qquad (\vec{v})(\Theta\circ\overline{D})^hx_h+...+(\vec{v})(\Theta\circ\overline{D})^1x_1+(\vec{v})x_0=\vec{0}$$
for some integer $h$ such that $1\leq h\leq n-1$, where $\overline{D}=\mathrm{diag}(M_1,...,M_r)$. By considering only non-trivial vectors $\vec{v}\in (\mathbb{F}_q^{\theta})^n$,
$(\#\#)$ becomes simply
$$(\#\#')\qquad \vec{v}\cdot (\overline{D}_hx_h+...+\overline{D}_1x_1+Ix_0)=\vec{0}$$
which immediately implies the existence of non-trivial solutions $x_h,...,x_1,x_0\in\mathbb{F}_q$ of the following equation
$$(\#\#'')\qquad\det (\overline{D}_hx_h+...+\overline{D}_1x_1+Ix_0)=0,$$
where $\overline{D}_i=\mathrm{diag}((M_1)_i,...,(M_r)_i)$ and $(M_j)_i$ is as in Lemma \ref{powers} for every $j=1,...,r$ and $i=1,...,h$.
Observe that $(\#\#'')$ is equivalent to
$$\det (\mathrm{diag}((M_1)_hx_h+...+(M_1)x_1+Ix_0,...,(M_r)_hx_h+...+(M_r)x_1+Ix_0))=$$
$$=\det ((M_1)_hx_h+...+(M_1)x_1+Ix_0)\cdot ...\cdot\det ((M_r)_hx_h+...+(M_r)x_1+Ix_0)=0,$$
i.e.
$$F(x_0,x_1,...,x_h):=F_{1,h}(x_0,x_1,...,x_h)\cdot ...\cdot F_{r,h}(x_0,x_1,...,x_h)=0,$$
where $F_{i,h}(x_0,x_1,...,x_h):=\det ((M_i)_hx_h+...+(M_i)x_1+Ix_0)$ for every $i=1,...,r$.
In this case, the zero locus $Z(F(x_0,x_1,...,x_h))$ of $F(x_0,x_1,...,x_h)$ on the projective space
$\mathbb{P}^{h}({\mathbb{F}_q})$ is a complete intersection of type $(d_1,...,d_r)$, where $d_i:=\deg F_{i,h}(x_0,x_1,...,x_h)$,
and its (rational) points are solutions of $(\#\#'')$.

\bigskip

When $\theta=id$, the above arguments work also with $M$ instead of $\Theta\circ A$
and give us a method to construct all the linear codes $\mathcal{C}\subseteq\mathbb{F}^n_q$
invariant under a matrix $M$ as in $(*)$. Indeed, for $h=n-1$ the equation $(\#'')$ becomes simply 
$$\det (M^{n-1}x_h+...+M^1x_1+Ix_0)=0.$$ 
Thus by any point $(x_0,x_1,...,x_{n-1})\in Z_{n-1,n}$ we can
construct a polynomial $p=p(X)\in\mathbb{F}_q[X]$ such that $\det p(M)=0$. 

\medskip

By the following Magma~\cite{M} program we can find all the solutions of the equation
$\det (M^{n-1}x_h+...+M^1x_1+Ix_0)=0$ in $\mathbb{F}_a[X] :$

\begin{verbatim}
F<w>:=GF(a);
P<x>:=PolynomialRing(F);
PointsCode := function(M);
 k:=Parent(M[1,1]); 
 n:=Nrows(M); 
 P<[x]>:=ProjectiveSpace(k,n); 
 X:=Scheme(P,Determinant(&+[x[i+1]*M^i : i in [0..n]])); 
 pts:=Points(X);
 ll:=[]; 
  for pp in pts do
   p:=Eltseq(pp);
   ll := ll cat [NullSpace(&+[p[i+1]*M^i : i in [0..n]])]; 
  end for; 
 return ll;
end function;
\end{verbatim}

\medskip

\noindent In fact, when $\theta=id$, we can say more about the above polynomial
$p\in\mathbb{F}_q[X]$.

\begin{prop}\label{remark fin}
Assume that $\theta=id$. Let $m\in\mathbb{F}_q[x]$ be the
minimal polynomial of an invertible matrix $M$. If
$g=gcd(p,m)$ for some polynomial
$p\in\mathbb{F}_q[X]$, then
\begin{enumerate}
\item $\mathrm{Ker\ }p(M)=\mathrm{Ker\ }g(M)$; \item $\mathrm{Ker\
}p(M)\neq\vec{0}\iff g\neq 1$.
\end{enumerate}
\end{prop}

\noindent\textit{Proof}. Note that $g=pa+mb$ for some
polynomials $a,b\in\mathbb{F}_q[X]$. Hence $g(M)=p(M)a(M)$ and
$p(M)=g(M)b(M)$. This shows that $\mathrm{Ker\
}p(M)\subseteq\mathrm{Ker\ }g(M)$ and $\mathrm{Ker\
}g(M)\subseteq\mathrm{Ker\ }p(M)$ respectively, i.e. $\mathrm{Ker\
}p(M)=\mathrm{Ker\ }g(M)$.

Finally, to prove $(2)$, first assume that $g=1$. Then
$g(M)=p(M)a(M)$ is the identity matrix. This shows that $\det
p(M)\cdot\det a(M)=1$, i.e. $\det p(M)\neq 0$, but this gives a
contradiction. On the other hand, if $g\neq 1$ then
$m=hg$ and $p=lg$ for some polynomials
$h,l\in\mathbb{F}_q[X]$. Hence $h(M)g(M)$ is the zero
matrix. Since $\deg h<\deg m$ and $m$ is the minimal
polynomial of $M$, we deduce that $\det g(M)=0$. Thus we get $\det
p(M)=\det (l(M)g(M))=0$, i.e. $\mathrm{Ker\ }p(M)\neq\vec{0}$.
\hfill $\square$

\bigskip

\section*{Conclusion}

\noindent In this paper we study the main properties of codes invariant by a semi-linear transformation of $\mathbb{F}_q^n$ for $n\geq 2$
in the non-commutative ring $\mathbb{F}_q[X,\theta]$, where $\theta :\mathbb{F}_q\to\mathbb{F}_q$
is an automorphism of the finite field $\mathbb{F}_q$ with $q$ elements. In particular, we introduce the notion
of product semi-linear codes and we study their Euclidean, quasi-Euclidean and Hermitian dual codes.
The main ingredient here is the construction of an injective map
which transforms a product semi-linear code into a skew cyclic code for taking advantages of the nice properties of these latter well-known codes.
In addition, we show some connections between these three types of dual codes and we give
an encoding and decoding algorithm with product semi-linear codes and a method to construct a code invariant under a semi-linear transformation of $\mathbb{F}_q^n$.
It is hopped that all these results will be a future topic of interest for further studies on generalized cyclic codes and skew quasi-cyclic codes.

\bigskip
\bigskip
\bigskip

\noindent\textbf{Acknowledgements.} The authors would thank Prof. A. Laface 
for useful remarks about the MAGMA programs and for some interesting discussions about the final form of this paper.


\newpage


\begin{thebibliography}{1}

\bibitem{M} W. Bosma, J. Cannon and C. Playoust, \emph{The
Magma algebra system. I. The user language}, J. Symbolic Comput.
{\bf 24} (1997), no. 3-4, 235--265.

\bibitem{BU1}
D. Boucher and F. Ulmer, \emph{A note on the dual codes of module
skew codes}, Cryptography and coding, Lecture Notes in Comput.
Sci. {\bf 7089}, Springer, Heidelberg, 2011, 230--243.

\bibitem{BU2}
D. Boucher and F. Ulmer, \emph{Linear codes using skew polynomials
with automorphisms and derivations}, Des. Codes Cryptogr. 2013,
{\bf DOI} 10.1007/s10623-012-9704-4.

\bibitem{BU3}
D. Boucher and F. Ulmer, \emph{Codes as modules over skew
polynomial rings}, Cryptography and coding, Lecture Notes in
Comput. Sci. {\bf 5921}, Springer, Berlin, 2009, 38--55.

\bibitem{HK}
M. Homma and S.J. Kim, \emph{An elementary bound for the number of
points of a hypersurface over a finite field}, Finite Fields Appl.
{\bf 20} (2013), 76--83.

\bibitem{HP}
W. Cary Huffman and V. Pless, \emph{Fundamentals of
error-correcting codes}, Cambridge University Press, Cambridge,
2003.

\bibitem{J}
N. Jacobson, \emph{Pseudo-linear transformations}, Ann. of Math.
(2) {\bf 38} (1937), no. 2, 484--507.

\bibitem{LL}
Zhuo-Jun Liu and Dong-Dai Lin, \emph{A class of generalized cyclic
codes}, Acta Math. Appl. Sinica (English Ser.) {\bf 16} (2000),
no. 1, 53--58.

\bibitem{LS}
S. Ling and P. Sol\'e, \emph{On the algebraic structure of
quasi-cyclic codes. I. Finite fields}, IEEE Trans. Inform. Theory
{\bf 47} (2001), no. 7, 2751--2760.

\bibitem{MacWS}
F.J. MacWilliams, N.J.A. Sloane, \emph{The Theory of Error-Correcting Codes},
North-Holland Publishing Company, Amsterdam, New York, Oxford, 1977.

\end{thebibliography}
\end{document}